%% file: GLANCE_MNRAS_final.tex
\documentclass[fleqn,usenatbib]{mnras}

\usepackage{newtxtext,newtxmath}

\usepackage[T1]{fontenc}

\DeclareRobustCommand{\VAN}[3]{#2}
\let\VANthebibliography\thebibliography
\def\thebibliography{\DeclareRobustCommand{\VAN}[3]{##3}\VANthebibliography}

\usepackage{graphicx}	
\usepackage{amsmath}	
\usepackage{appendix}
\usepackage{caption}
\usepackage{cuted}
\usepackage{placeins}  
\usepackage{array}
\usepackage{txfonts}
\usepackage{longtable,ltcaption}
\usepackage{lscape}
\usepackage{natbib}
\usepackage{lscape}
\usepackage{amsmath}
\usepackage{wasysym}
\usepackage{graphics}
\usepackage{txfonts}
\usepackage{color}
\usepackage{times}
\usepackage{parskip}
\usepackage{pdflscape}
\usepackage{geometry}
\usepackage{marginnote}
\usepackage[table]{xcolor}
\usepackage{tabu}
\usepackage{stfloats}
\usepackage{mathtools} 
\usepackage{nicefrac}
\usepackage{float}
\usepackage[normalem]{ulem}
 
\usepackage{enumitem}

\newcommand{\object}[1]{\textrm{#1}}
\def\exG{\object{NGC 1566}}
\def\NGC1566{NGC1566}

\def\glance{{\sc Glance}}
\def\starlight{{\sc Starlight}}
\def\fado{{\sc Fado}}
\def\dyn{{\sc Dynamite}}
\def\kin{{\sc kinemetry}}
\def\ppxf{{\sc pPXF}}
\def\bayes{{\sc Bayes-LOSVD}}
\def\pegase{{\sc Pegase}}
\def\bang{{\sc BANG}}
\def\ry{${\cal R}$emove${\cal Y}$oung}
\def\RY{${\cal RY}$}

\def\gist{{\sc GIST}}
\def\gandalf{{\sc GandALF}}
\def\pipe3d{{\sc Pipe3D}}
\def\fit3d{{\sc FIT3D}}

\def\D4000{$D_{4000}$}

\def\mlr{$\Upsilon_{\star}$}

\newfont{\nlx}{cmssdc10 scaled 900}
\newfont{\mfont}{cmssdc10 scaled 760}
\definecolor{myblue1}{rgb}{0.0,0.604,0.831} 
\definecolor{myblue2}{rgb}{0.0,0.49,0.6745}
\definecolor{myblue3}{rgb}{0.0156,0.4078,0.9921}
\definecolor{myblue4}{rgb}{0.0,0.44,0.87}
\definecolor{myred1}{rgb}{0.529,0.019,0.017}
\definecolor{mycyan}{rgb}{0.63921569,0.0,0.48235294}

\title[]{\glance: A Comprehensive Framework for Galactic
Archaeology}

\author[I. Breda et al.]{Iris Breda$^{1}$\thanks{E-mail: iris.breda@univie.ac.at},
Glenn van de Ven$^{1}$,
Sabine Thater$^{1}$,
Federica Mauro$^{1}$,
Stergios Amarantidis$^{2}$,
J. Falc\'on-Barroso$^{3,4}$,
\newauthor
Prashin Jethwa$^{1}$,
Anja Feldmeier-Krause$^{1}$,
Masato Onodera$^{5,6}$,
Gerhard Hensler$^{1}$ \&
Gauri Sharma$^{7}$
\\
$^{1}$Department of Astrophysics, University of Vienna, Türkenschanzstraße 17, 1180 Vienna, Austria\\
$^{2}$Institut de Radioastronomie Millim\'etrique (IRAM), Avenida Divina Pastora 7, Local 20, E-18012, Granada, Spain\\
$^{3}$Instituto de Astrof\'isica de Canarias, Calle V\'ia L\'actea s/n, E-38205, La Laguna, Tenerife, Spain\\
$^{4}$Dep. de Astrof\'isica, Universidad de La Laguna, Av. del Astrof\'isico Francisco S\'anchez s/n, E-38206, La Laguna, Tenerife, Spain\\
$^{5}$Graduate Institute for Advanced Studies, SOKENDAI, 2-21-1 Osawa, Mitaka, Tokyo 181-8588, Japan\\
$^{6}$Subaru Telescope, National Astronomical Observatory of Japan, 650 N Aohoku Pl, Hilo, HI96720, Japan\\
$^{7}$University of Strasbourg, CNRS UMR 7550, Observatoire Astronomique de Strasbourg, F-67000 Strasbourg, France
}

\date{Accepted XXX. Received YYY; in original form ZZZ}

\pubyear{2025}

\begin{document}
\label{firstpage}
\pagerange{\pageref{firstpage}--\pageref{lastpage}}
\maketitle

\begin{abstract}
A central topic in extragalactic astronomy is understanding the formation and evolutionary histories of galaxies. These systems often comprise multiple structural components with distinct physical and dynamical properties, making it challenging to disentangle their individual contributions. Aiming at investigating the true structure of the inner stellar disk, we have developed a comprehensive pipeline for the chronochemical and dynamical analysis of galaxies (\glance: Galactic archaeoLogy via chronochemicAl \& dyNamiCal modElling). The presented pipeline employs several state-of-the-art techniques by integrating them into a single, automated pipeline, enabling streamlined analysis of integral-field spectroscopy data, by allowing users to easily and directly extract valuable information on stellar populations, kinematics, dynamics, and gas properties. It automates multiple analysis techniques, including stellar population synthesis (\fado, \starlight, post-processing with \ry, kinematic extraction (\ppxf, \bayes), and dynamical modelling (\dyn). It handles tasks such as Galactic extinction correction, de-redshifting, Voronoi binning, and nebular continuum correction, while offering extensive customization options. Parallel processing significantly reduces computational time. When applied to MUSE data sampling the central region of \exG, this methodology reveals that its stellar disk significantly deviates from the conventional exponential model, challenging the assumption of universality in disk morphology. In summary, this work presents a powerful, publicly available pipeline for conducting galactic archaeology, designed to advance our understanding of the formation and evolution of galaxies.\\
\end{abstract}




\section{Introduction \label{intro}}

Understanding the formation and evolutionary histories of galaxies is a central challenge in extragalactic astronomy. Galaxies are complex systems composed of multiple structural components (e.g. bulges, disks, bars, halos), each with distinct stellar populations and kinematics, and disentangling such stellar structures remains a complex task. In the advent of integral-field spectroscopy (IFS), which supplies spatially resolved spectral data of a significant volume of galaxies near and far, such challenge has become attainable. Modern IFS surveys and instruments (e.g. CALIFA, MaNGA, SAMI, and Multi Unit Spectroscopic Explorer, MUSE; \citealp{califa,manga,sami,muse}) offer spatially resolved spectra for thousands of galaxies. To leverage these rich datasets, the community has developed powerful analysis pipelines that combine multiple techniques into unified frameworks. For instance, the \pipe3d\ pipeline \citep{pipe3d} uses full spectral fitting (via the \fit3d\ code; \citealt{fit3d}) to derive both stellar population properties and ionized gas characteristics from IFS datacubes, yielding consistent data products for surveys like CALIFA, MaNGA and SAMI. Similarly, the \gist\ pipeline \citep[Galaxy IFU Spectroscopy Tool,][]{gist} offers an all-in-one toolbox -- from data preparation to science analysis to visualization -- that extracts stellar kinematics, performs emission-line diagnostics, and derives stellar population ages/metallicities from IFS data by means of well-established routines (e.g. \ppxf; \citealp{CapEms04,Cap17} for kinematics and star-formation histories, SFHs; \gandalf; \citealt{gandalf} for emission-line fitting). These pipelines are modular by design, allowing users to easily adapt them to different datasets and scientific goals, exemplifying how streamlined, reproducible analysis workflows can maximize the scientific return of modern spectroscopic surveys by enabling astronomers to efficiently map galaxies’ stellar content and kinematics.

While existing IFS pipelines focus on extracting stellar populations and kinematics, they generally do not incorporate the next analytical step: building dynamical models of galaxies that can disentangle the mass and orbital distributions of distinct components. In particular, to truly disentangle and characterize the different stellar components that populate galaxies, one must adopt a comprehensive approach in order to extract as much information as possible from these rich data-sets, including orbital architectures.

The Schwarzschild orbit-superposition method \citep{Sch79} provides a powerful tool for understanding the dynamics of galaxies. Elementary, this method involves simulating stellar motions within a galaxy, by integrating orbits in a gravitational potential, and assigning weights to these orbits to match the observed stellar mass distribution and kinematics. Galaxy dynamics are modelled by integrating orbits within a complex gravitational potential dictated by the stellar and dark matter components, as well as a central super-massive black hole. This allows for assessing the relative contribution of specific orbital structures drawn from an orbit library, while determining best-fit parameters, such as central black hole mass (M$_{\bullet}$), stellar mass-to-light ratio (\mlr), the morphology of the different kinematic components, and dark matter halo properties. Different levels of symmetry have been implemented in the Schwarzschild method \citep[e.g.,][]{Cap06,VanBos08,VasAth15,VasVal20,Neu21}. This method has been utilized for estimating supermassive black holes masses \citep[e.g.,][]{Kra09,Tha17,Tha19,Lie20,Que21}, analysing the internal orbital structures of stellar clusters \citep[e.g.,][]{Van06,Fel17,Fah19}, investigating early-type galaxies \citep[e.g.,][]{Cap06,Fah19,Poci19,Jin20,dBro21,Tha23}, and recently extended to galaxies more complex structures \citep[e.g.,][]{Zhu18b,Zhu18c,LipTho21}. Additionally, it has been instrumental in identifying different dynamical components within stellar systems \citep[e.g.,][]{Van06,Lyu13,BreHel14,Kra15,Jin24}.

Here, we present a comprehensive pipeline designed to perform galactic archaeology by analysing spatially resolved spectral data, streamlining the photometric, chronochemical, and dynamical analysis of galaxies. With the increasing availability of IFS data, a tool that integrates multiple analysis techniques into a unified, automated framework -- enabling the community to extract extensive information with minimal effort -- becomes increasingly essential. Although this tool was developed to re‑examine the underlying assumptions of photometric decomposition -- especially the notion that the stellar disk retains an exponential profile into the galactic centre -- and to disentangle the overlapping contributions of bulge and disk, the pipeline's modular and highly customizable design allows for flexible application, enabling users to tailor analyses to their specific research needs. This methodology can be applied to investigate, e.g., the spatial progression of star formation and chemical enrichment, to assess the dynamical nature of different components, or to connect observed present-day structures to their formation epochs. Additionally, it can be applied to test key assumptions in structural decomposition (e.g., the exponential nature of disks), quantify the impact of non-axisymmetric features on galaxy dynamics, and refine measurements of fundamental parameters such as stellar mass-to-light ratios and central black hole masses.

This article is structured as follows: Section~\ref{meth} outlines the adopted methodology, Section~\ref{res} presents the key outputs from applying the pipeline to the innermost region of \exG, and Section~\ref{conc} summarizes the main conclusions.
\section{Methodology}\label{meth}

To effectively analyse the available high-quality MUSE data, a comprehensive pipeline implemented in \texttt{Python} was developed, integrating several modules designed to perform spectral analysis, with a central reference radius being defined by surface photometry. Key techniques implemented include Voronoi binning \citep{voronoi}, stellar population synthesis (SPS) using \fado\ \citep{fado} and/or \starlight\ \citep{starlight}, 
emission line fitting, extraction of stellar kinematics employing \ppxf\ \citep{CapEms04,Cap17} and/or \bayes\ \citep{FalMar21}, and dynamical modelling with \dyn\ \citep{VanBos08,Jet20,dyn}. 
The pipeline also incorporates photometric fitting to identify a central reference radius, facilitating the comparison of physical properties between inner and outer regions. 
Its modular nature allows for the flexible selection of individual techniques to deploy in each run, ensuring adaptability to the specific requirements of the analysis. As input, the pipeline simply requires an ASCII table specifying the galaxy name, the minimum and target signal-to-noise ratio (SNR) for Voronoi tessellation, and the literature super-massive black hole (M$_{\bullet}$) mass to be assumed for dynamical modelling. Note that the user must configure certain settings before using the pipeline for the first time, such as specifying paths to SSP models, external executables (e.g., \fado, \starlight), and additional parameters including preferred methods for specific routines, and other analysis options. For a complete list, we refer to the pipeline’s documentation/source code.
As this tool is publicly available, we encourage the community to actively contribute to its ongoing development, fostering a collaborative and inclusive effort\footnote{The pipeline and its documentation are available at \url{https://gitlab.com/iris.b/glance}.}.

\subsection{Photometric Analysis}\label{meth-phot}

Structural decomposition was achieved by performing surface photometry on $K$-band data. The near infrared (NIR) primarily reveals the underlying old stellar component, i.e., the $\textit{galactic skeleton}$ that contains the bulk of the stellar mass of the galaxy. Opportunely, the canonical AGN power-law also displays its minimum at these wavelengths, translating into a significant reduction of the AGN luminosity in the NIR when compared to the optical \citep{Pad17}, while also being less affected by dust attenuation. Therefore, by utilizing $K$-band data instead of optical frames, one can effectively mitigate the dominant emission from optically-bright young stellar populations, whose contribution to the overall stellar mass is marginal, alongside potential AGN signatures, thus uncovering the primordial disk, i.e., an old stellar disk $\textit{skeleton}$, most probably formed in parallel with the majority of the bulge's (and bar, if existing) stellar mass. $K$-band observations are manually retrieved from the repositories of VISTA (VHS, \citealt{vhs}, or VIKING, \citealt{viking}) and UKIRT (UKIDSS, \citealt{ukidss}, or UHS, \citealt{uhs}). To note that, if desired, instead of using $K$-band imaging, the pipeline includes routines to automatically download alternative photometric data such as HST or SDSS images.

Prior to surface photometry, this module processes the photometric data by clipping foreground stars either automatically or manually (if the user is unsatisfied with the automatic procedure, they are prompted to perform the clipping manually). In the automatic mode, compact bright sources are detected above a $10\sigma$ threshold, the largest source -- assumed to be the galaxy -- is excluded, and the remaining masks are slightly dilated to ensure complete removal of stellar flux. The resulting masked image is then used for modelling and subtracting the sky background. The workflow is predominantly automatic, but allows targeted user intervention at key stages (validation of the sky model, definition of the galaxy aperture, and PSF/stellar selection) whenever the resulting outcome requires refinement. The surface brightness profile (SBP) is then extracted either from elliptical isophote fitting with the \textsc{photutils.isophote} package or, alternatively, from MGE-based SBPs computed with the \texttt{sbProf} utility. The SBP, stored in calibrated surface-brightness units, is corrected for Galactic extinction using the extinction values returned by NED/IRSA together with the photometric zero point in the image header. GLANCE subsequently performs an interactive 1D decomposition of the SBP, in which the user selects radial ranges representative of the disc, bar (when present), and bulge; these segments are fitted with exponentials and PSF-convolved S\'ersic profiles, following the approach by \citealt{BP18}, to derive structural parameters and characteristic radii (e.g. the 24~mag~arcsec$^{-2}$ isophotal radius of the disc). The resulting 1D best-fitting parameters are finally used as informed initial conditions for a fully 2D multi-component fit with \textsc{galfit}, which yields a PSF-convolved model of the galaxy (disc, bulge, bar and sky) and the corresponding residual maps.

\subsection{Spectral Analysis}\label{meth-spec}

The pipeline's spectral workflow is designed to extract detailed information of the sampled galaxies, including the characterization of their stellar populations, emission line analysis, and kinematical extraction and subsequent dynamical modelling. The following presents a concise description of each phase of the data processing:

$\bullet$ \uline{Data preparation}: MUSE data cubes are de-redshifted and corrected for Galactic extinction. Herein, the user might request for Voronoi binning \citep{voronoi}, defining the target SNR estimated at the emission-line-free continuum between 6390 and 6490 \AA. If desired, the user might also mask possible foreground stars and to select a segment of the data-cube. The pre-processed spectra (i.e., binned, de-redshifted and corrected for Galactic extinction) are therefore ready to be analysed in the subsequent steps.

$\bullet$ \uline{Spectral synthesis with \fado}: \fado\ \citep{fado} is a SPS tool that uniquely accounts for both stellar and nebular emissions. This self-consistent approach is particularly important for galactic regions with active star formation, where nebular emission significantly contaminates the observed stellar spectrum. As a result, \fado\ offers more accurate determinations of star formation histories (SFHs), ensuring that contributions from ionized gas are properly considered. Spectral fits in the (rest-frame) range between 4730 $\AA$ and 8740 $\AA$ were carried out, adopting two simple stellar population (SSP) libraries based on two sets of models: BC03 \citep{BruCha03} and CB19 \citep{BruCha03,Plat19}. The default SSP library based on the BC03 models comprises 152 SSPs for 38 ages between 1 Myr and 13 Gyr for four stellar metallicities (0.05, 0.2, 0.4 and 1.0 $Z_{\odot}$), referring to a Salpeter initial mass function \citep{Sal55} and Padova 2000 tracks. As for the CB19 models, the default SSP library includes 
176 SSPs for 44 ages between 0.1 Myr and 13.5 Gyr and four stellar metallicities (0.05, 0.2, 0.4 and 0.9 $Z_{\odot}$), referring to a Salpeter initial mass function and PARSEC \citep{Bre12,Chen15}, and TP-AGB \citep{Mar13} evolutionary tracks. 
Processing with \fado\ is conducted in parallel, maximizing computational efficiency by distributing the workload across all available CPUs. The pipeline architecture facilitates straightforward customization, enabling the inclusion of alternative model sets or preferred libraries (e.g., SSPs with super-solar metallicities). If desired, Monte Carlo (MC) simulations (default being 100 realizations) can be performed by adding random noise at the observational level to the spectra enabling the derivation of realistic uncertainties for the derived SFHs, mean stellar ages and metallicities, and mass estimates.

$\bullet$ \uline{Estimation and subtraction of the nebular continuum}: 
If the user desires, the nebular continuum contribution might be empirically estimated (Papaderos and {\"O}stlin, 2025, submitted for publication in A\&A) and subtracted from the observed binned spectra prior to fitting with \starlight, \ppxf, and \bayes, thus decontaminating the spectra from its potential nebular emission in all subsequent steps of the analysis (this operation will not be applied for \fado, since the latter already performs it by default). This is accomplished by assessing the observed nebular continuum and scaling its canonical form to the number of produced Lyman continuum photons (being equivalent to the H$\alpha$ flux, under the assumption of case-B recombination). This correction is particularly important in starburst galaxies, where the nebular continuum can represent a significant fraction of the total optical emission and strongly affect the derived stellar population properties \citep{Mir25}. The procedure goes as follows:
\begin{enumerate}
\item Estimate the V-band extinction in the nebular component \( A_V^{\rm neb} \) using the Balmer decrement:\\ \( A_V^{\rm neb} \) = 7.96 log((H$\alpha$/H$\beta$)$_{\rm obs}$ / (H$\alpha$/H$\beta$)$_{\rm theo}$),\\
with (H$\alpha$/H$\beta$)$_{\rm theo}$ = 2.86;
\item Compute the dimming of the nebular component for the previously derived \( A_V^{\rm neb} \), under the adopted extinction law. Specifically, the dimming is calculated as:\\
Neb$_{\rm obs}$ = 10$^{\rm 0.4 A_V^{\rm neb} \cdot\ C}$,\\
with C being the \cite{Cal00} extinction function, with a ratio of total to selective extinction R$_{\rm V}$ of 3.1;
\item Apply \( A_V^{\rm neb} \) to the pre-computed UV-through-IR (canonical) SED of the nebular continuum Neb$_{\rm C,ref}$:\\
 Neb$_{\rm C,obs}$ = Neb$_{\rm C,ref}$ / Neb$_{\rm obs}$;
\item Scale the extincted nebular continuum SED to the observed H$\alpha$ flux, ensuring that the nebular continuum level at 6563 \AA\ is \( 2.03 \times 10^{-4} \cdot {\rm H}\alpha_{\rm obs} \) (the scaling factor $s$ = $2.03 \times 10^{-4}$ was empirically determined through the atomic physics routine in the spectral evolution model \pegase, \citealt{pegase}). Therefore, the Neb$_{\rm C,obs}$ must be scaled by the factor $\psi$ = $s/f$, where \( f \) is the flux of the extincted nebular continuum Neb$_{\rm C,obs}$ at 6563 \AA;
\item The corrected spectrum can subsequently be obtained by subtracting $\psi \cdot\ {\rm Neb}_{\rm C,ref}$ to the spectrum corresponding to each bin.
\end{enumerate}





$\bullet$ \uline{Spectral synthesis with \starlight}: \starlight\ \citep{starlight} is a SPS code designed to interpret the observed spectra of galaxies by decomposing them into a combination of SSPs, allowing for the extraction of star formation histories, mean stellar ages, and metallicities. This module follows the same approach as described earlier for \fado, but with the inclusion of the MILES SSP library \citep{MILES_SSP} alongside BC03 \& CB19. The upper fitting limit for MILES is 7400 $\AA$. Similar to the \fado\ module, processing with \starlight\ is executed in parallel, optionally supporting MC simulations to ensure robust error estimation. 

It follows a brief description of the generated data-products for both \fado\ and \starlight, which consist of \texttt{fits} files comprising the maps of various physical quantities, figures displaying some of these maps (see Fig.~\ref{fado0}), and \texttt{pickle}\footnote{\texttt{pickle} files are \texttt{Python}'s serialized data format, storing objects (variables, models, etc.) in a binary form for efficient saving/reloading while preserving their original structure. Herein, the produced \texttt{pickle} (*.pkl) files store detailed per-bin results and metadata for reproducibility and further analysis.} files \citep{pkl} comprising detailed information on each fit. If the user wishes to derive realistic errors via MC simulations, additional images, and fits and \texttt{pickle} files are produced. The \texttt{fits} files contain 2D maps of key physical parameters\footnote{Metadata in the header provides details on the setup used, such as the fitting range, SSP library used, calibration details, and flux conversion instructions.}. These include stellar properties -- mass- and luminosity-weighted ages (Gyr) and metallicities (Z$_\odot$), present-day and total ever-formed stellar masses (logM$_\odot$), and stellar surface density -- emission-line fluxes ($10^{-17}$ $\cdot$ erg s$^{-1}$ cm$^{-2}$) and EWs (\AA) for several emission lines -- and in the case of \fado, nebular (E(B-V)$_{\text{neb}}$) and stellar ($A_{\rm V}$) extinction maps. In addition, composite \texttt{png} images featuring stellar population properties and key emission-line fluxes/EWs are generated, as well as an additional figure displaying emission-line diagnostics \citep[BPT,][]{BPT} and the Kennicutt relation \citep{ken}.

Note that 1) while the files from \fado\ and \starlight\ share a similar structure for key properties, they may differ in the in/exclusion of secondary physical quantities; 2) similar files are generated during MC simulations, including posterior distributions and uncertainty estimates of quantities obtained by SPS; 3) while this section highlights the key outputs, the files contain additional details. For a comprehensive description of their contents, refer to the pipeline's source code.



$\bullet$ \uline{Emission-line analysis}: After subtracting the best-fitting stellar continuum obtained with \starlight, fluxes and equivalent-widths of several emission lines (i.e., H$\beta$, OIII$_{\rm a,b}$, NII$_{\rm a,b}$, H$\alpha$, SII$_{\rm a,b}$) are determined by employing the \texttt{Python} package \textsc{PySpeckit} \citep{pyspeckit1,pyspeckit2} in parallel. To account for dust attenuation within the galaxy, emission line fluxes are subsequently corrected for internal extinction, adopting the \cite{G23} reddening law, and using the Balmer decrement derived from the observed ratio of hydrogen recombination lines (i.e., H$\alpha$/H$\beta$).

$\bullet$ \uline{Post-processing with \ry}: \ry\ \citep[\RY,][]{RY} is a tool designed to isolate and analyse the older stellar components in galaxies by removing the contributions from younger stellar populations (below a user-defined age cutoff). The tool operates by post-processing the output from SPS models applied to IFS data cubes. By specifying an age threshold, \RY\ computes the spectral energy distribution, surface brightness, and stellar mass density distribution of stellar populations older than the selected cutoff. By default, this module applies \RY\ to the output of \fado\ and \starlight\ for eight age cuts: 31, 102, and 302 Myr, as well as 1.02, 3.01, 5.01, 7.01, and 9.01 Gyr. 

$\bullet$ \uline{Kinematics extraction with \ppxf}: \ppxf\ \citep{CapEms04,Cap17} is a versatile tool for extracting kinematics and  stellar population properties from galaxy spectra through full-spectrum fitting. It employs a maximum penalized likelihood approach to fit the observed spectra with a library of stellar spectra, allowing for the determination the line-of-sight velocity distribution (LOSVD) of stars (and gas by annexing Gaussian emission‐line templates to the library), stellar ages, and metallicities. The nebular continuum free spectra is fed to \ppxf\ to extract kinematical maps for 4  Gaussian-Hermite (GH) moments, for two sets of templates (IUS, \citealt{IUS}, and MILES, \citealt{MILES}). Stellar kinematics are assessed in 2 ways -- by extracting purely stellar kinematics while masking emission-lines, and stellar + gas kinematics. In the latter, the gas fit can be configured via a flag, having the following options: 1) all lines share the same kinematics, with both a narrow and broad component; 2) each emission-line is treated as its own (narrow) component; 3) lines are grouped by species (e.g., Balmer lines share one LOSVD, each forbidden doublet another), sharing their kinematics within each group (default). In either case, the stellar templates are convolved with the MUSE spectral resolution function, and the fitting range can be defined as 1) the highest wavelength range shared by both observations and templates (default); 2) truncated to the blue spectral range -- 4820 to 5750 $\AA$; 3) truncated to the red spectral range -- 8400 to 8720 $\AA$. 
 As previously, kinematic modelling is performed in parallel, being supplemented with MC simulations. 
Analogous to the spectral synthesis module, the kinematic module produces: 1) \texttt{fits} files containing 2D maps of the mean velocity, velocity dispersion, and higher-order velocity moments, along with metadata, such as the best-fit kinematic position angle (PA) and systemic velocity; 2) diagnostic images displaying the kinematic maps (see Fig.~\ref{ppxf}); 3) \texttt{pickle} files containing detailed information on the fit, including the values of the kinematic quantities per bin and their associated errors. These files provide a comprehensive record of the fitting process, enabling further analysis and reproducibility. 
Upon completion, the systemic velocity and the kinematic PA are determined using the \texttt{Python} package \textsc{PaFit} \citep{pafit}. Subsequently, the systemic velocity is subtracted from the rotation velocity, and the median of each odd moment is subtracted from its corresponding value. This is done because the rotation velocity (and other odd moments) should be zero at the galaxy center; however, small offsets can arise from imperfect kinematic fitting. These offsets must be corrected before feeding the kinematic data to \dyn\ to ensure consistency in the dynamical modelling. Note that the modular design of the pipeline allows for relatively straightforward adaptation to support additional functionalities, such as population spectral synthesis via \ppxf.

$\bullet$ \uline{Stellar kinematics extraction with \bayes}: If desired, stellar kinematics can be extracted using the alternative method \bayes\ \citep{FalMar21}, which offers a Bayesian framework to analyse the LOSVD. This method employs Bayesian inference to obtain robust LOSVDs along with their associated uncertainties. It utilizes principal component analysis to reduce the dimensionality of the template base required for extraction, thereby enhancing computational performance. Additionally, the framework offers various regularization options to ensure reliable solutions. Comparably to the \ppxf\ module, the fitting range can be defined to cover the entire spectrum (though this is computationally intensive) or restricted to either the blue or red spectral range for faster processing. By default, IUS stellar templates are selected and no regularization is adopted. 


$\bullet$ \uline{Preparation for \dyn}: Dynamical modelling with \dyn\ requires a series of preparatory steps. Therefore, the pipeline has a module which ensures that every requirement for \dyn\ is met, such as the determination of the surface brightness distribution by parametrizing it as a Multi-Gaussian Expansion (MGE), and to create the necessary configuration files condensing all the information \dyn\ requires. With respect to the MGE, we determine it by utilizing the \texttt{Python} package \textsc{MgeFit} \citep{Cap02} on high-resolution photometric data (optimally HST, filters F814W or F606W). In case there is no available HST data, the pipeline contains an extra routine which allows to construct a combined photometric data-frame, using SDSS $r$ band imaging for the outskirts and a bandpass image obtained by convolving the MUSE data cube with the SDSS $r$ band filter transmission curve in the center. In this case, the MUSE-derived bandpass image is scaled to match the observed SDSS flux within the central region, to ensure consistency in integrated flux. Note that an MGE can be obtained from any other photometric dataframe the user desires, with its implementation being straightforward. \dyn\ additionally requires an accurate estimate of the MUSE PSF. In case there is available HST data, the MUSE PSF is estimated by convolving the higher-resolution HST image with progressively broader PSFs until it matches the MUSE image, using the \texttt{Python} package \textsc{imphot}\footnote{\url{https://imphot.readthedocs.io/}}. Alternatively, the PSF value from the MUSE header is assumed. The required input files for \dyn\ are generated, including stellar kinematics (containing information for each bin), aperture, bins, and configuration files\footnote{Visit \url{https://dynamics.univie.ac.at/dynamite\textunderscore docs/index.html} for detailed information on \dyn\ and its data preparation.}. Deviant points in the velocity field are identified using \kin\ \citep{pafit}, which fits the rotation velocity profile. Points that significantly deviate from this profile are assigned large uncertainties. The final kinematics file for \dyn\ includes these cleaned kinematic measurements, although an option exists to retain all data without modification. In addition, \dyn\ offers the possibility to include not only a light MGE (as derived from photometric data) but also a mass MGE. Thus, the pipeline provides a routine to estimate the mass MGE based on the stellar mass estimates obtained trough spectral synthesis in previous steps. If this is requested, \mlr\ will be estimated within each Gaussian, which will be translated to a mass MGE by multiplying the luminosity within each Gaussian by the respective \mlr. Note that, at this stage of development, we avoid extrapolating the mass MGE beyond the MUSE field-of-view (FoV). This is relevant only when the FoV does not encompass the entire galaxy, in which case the mass MGE is truncated at the FoV limit to ensure that mass estimates rely solely on the available spectral data. Diagnostic plots are produced to assess the quality of the data and inputs. Minimal user input may be required while running this module.

$\bullet$ \uline{Dynamical modelling with \dyn}: \dyn\ \citep[][and references therein]{dyn} is a software package designed for dynamical modelling of galaxies and globular clusters. It utilizes stellar kinematics and photometry to construct models that describe the mass distribution and dynamical state of these stellar systems. By fitting the optimal combination of orbits from precomputed orbit libraries to the observed data, \dyn\ enables the simultaneous characterization of the gravitational potential, dark matter content, and orbital structures within galaxies, accounting for the contributions of the black hole, dark matter halo, stellar component, and \mlr. 
The default orbit library employs a grid of 21 $\times$ 10 $\times$ 7 in the three integrals of motion (energy $E$, and two additional integrals $I2$ and $I3$), for both tube and box orbits. Each grid point is further refined with a dithering factor of 5$^3$, resulting in a total of 3 $\times$ 21 $\times$ 10 $\times$ 7 $\times$ 5$^3$ = 110,250 orbits. This configuration is well-suited for high-quality datasets, typically resulting in adequate models.



$\bullet$ \uline{Orbital decomposition}: Decomposition of the orbits into their cold, warm and hot components is performed by utilizing the function \textsc{analysis} from \dyn\ \citep[see][]{San22}. The limiting values of circularity ($\lambda_{\rm z}$) adopted to define cold, warm and hot components was set to the default values, 0.8, 0.25, and -0.25 (i.e., cold orbits are defined as having $\lambda_{\rm z}$ > 0.8, warm orbits as 0.25 < $\lambda_{\rm z}$ < 0.8, hot orbits as -0.25 < $\lambda_{\rm z}$ < 0.25, and counter-rotating orbits as $\lambda_{\rm z}$ < -0.25). Uncertainties on the SBPs of the different dynamical components are derived by accounting for all models whose $\chi^2$ values lie within 1$\sigma$ of the best-fit solution. To note that the estimated profiles are not typical single-band SBPs, but are obtained by converting orbit-based mass fractions into light using the best-fitting mass-to-light ratio and the total MUSE luminosity.

In addition, the pipeline supports the incorporation of SINFONI data, allowing users to include kinematics extracted in the near-infrared in their analysis.

\subsection{Perspectives for Further Development}\label{meth-future}

Since the pipeline is designed as a flexible and evolving framework, certain modules still require further development, while additional functionalities are foreseen for future implementation. Future developments will aim to extend this integration even further, for example by incorporating orbit colouring and additional diagnostic modules, progressively building a more complete framework for galactic archaeology. Modules that are planned to be incorporated in future versions of \glance\ include:

$\bullet$ \uline{Dynamical modelling with \bang}: \bang\
 \citep{bang1,bang2} is a GPU-based code designed for the automated morpho-kinematic decomposition of galaxies. It employs analytical potential-density pairs to model galactic components, enabling efficient and reliable fits of both morphological and stellar kinematic properties. This approach facilitates the simultaneous analysis of galaxy photometry and kinematics, providing insights into the structural components of galaxies. Due to time constraints, this module is not fully completed or tested. However, it can be easily implemented in future versions of the pipeline or by users interested in contributing to its further development.
 
$\bullet$ \uline{Extrapolating the mass MGE beyond the FoV}: An additional feature envisioned for future versions of the pipeline is the extrapolation of the mass MGE beyond the FoV for galaxies where the available IFS data does not cover the entire system. While the current implementation deliberately truncates the mass MGE at the FoV limit to avoid relying on unconstrained assumptions, extending it beyond this region would allow for more complete dynamical modelling, specifically for extended galaxies. To achieve this, one possible approach is to correlate the mass estimates derived within the IFS FoV with the photometric data-frame used to extract the light MGE, anchoring the mass distribution to the larger-scale light profile. Alternatively, color information from different photometric filters can be used to estimate mass-to-light ratio variations outside the spectroscopic coverage, providing a basis for extrapolating the mass distribution in a physically motivated way.
 
$\bullet$ \uline{Orbit colouring}: The orbit colouring functionality in \dyn, which enables the classification and visualisation of orbits according to their properties, is still under development. Once this feature is fully implemented in \dyn, the pipeline will incorporate a dedicated module that automatically runs the new version and processes its output. This will allow users not only to fit observational data but also to analyse and interpret the internal orbital structure of galaxies in a more intuitive and physically meaningful way, further enhancing the pipeline’s diagnostic capabilities.

$\bullet$ \uline{Bar component modelling}: While the latest version of \dyn\ already supports the inclusion of a bar component in the gravitational potential, this functionality is not yet fully integrated into the pipeline. Future versions will provide the necessary tools to prepare input data for barred galaxy modelling, including, for instance, the extraction of a separate bar MGE. The pipeline will offer users the option to consider bar potentials when running \dyn, extending its applicability to barred systems.

$\bullet$ \uline{AI-driven analysis module (\glance+)}: Given the large volume and complexity of the outputs generated by the pipeline -- ranging from spatially resolved star formation histories obtained with different SPS codes and libraries, to orbital structures and dynamical parameters -- manual inspection becomes increasingly difficult. To address this, the development of an AI-driven analysis module is envisioned for future versions of the pipeline. This module will apply machine learning techniques such as representation learning and manifold learning to extract latent structures from \glance's high-dimensional outputs to uncover physically meaningful patterns, correlations, and groupings that may not be easily identified through traditional methods. Such functionality would provide users with an efficient way to extract new insights from large galaxy samples by cross-correlating all of \glance's different outputs.

\vspace{-2em}
\subsection{Basic Usage Instructions}

\begin{figure*}
    \centering
    \includegraphics[width=0.9\linewidth]{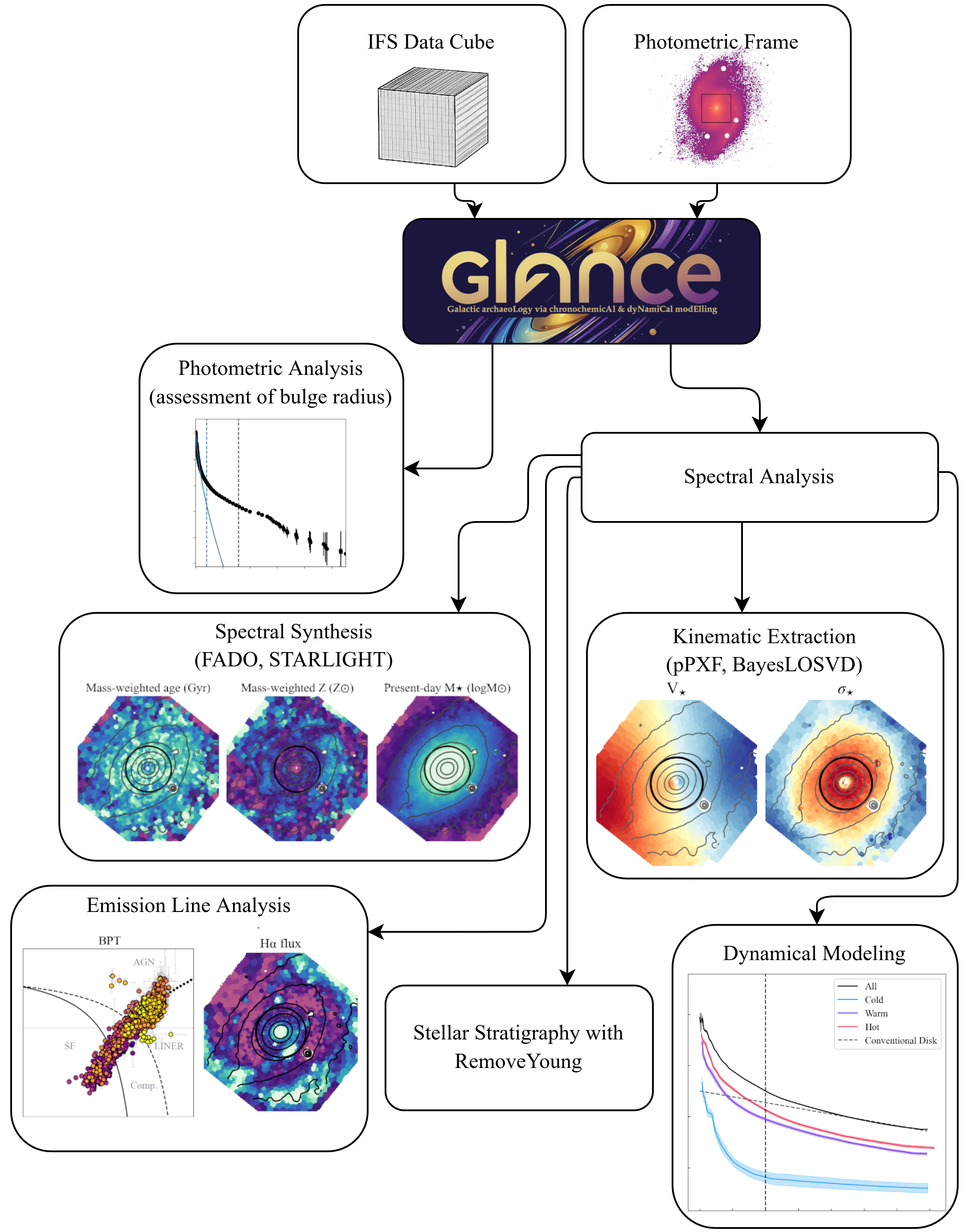}
    \caption{Schematic overview of the \glance\ pipeline. The code utilizes IFS data cubes and broad-band photometric frames, performing: photometric analysis, spectral synthesis via \fado\ \& \starlight, stellar stratigraphy trough \RY, emission-line analysis, stellar kinematic extraction via \ppxf\ \& \bayes, and dynamical modelling by means of \dyn. Each module can be executed independently, facilitating specific user requirements.}
    \label{scheme}
\end{figure*}

The \glance\ pipeline comprises several dedicated modules, each tailored to a specific component of the analysis (see Fig.~\ref{scheme}). Prior of running the pipeline for the first time, basic information must be provided through the interactive script \verb|GLANCE_run_1st_time.py|. This script reads a plain-text file (e.g. \verb|gal_info.txt|) containing the list of galaxy identifiers (e.g., \verb|NGC1566|; one \verb|galaxy_name| per line), assuming that the corresponding IFU data cubes follow the naming convention \verb|galaxy_name_DC.fits|. After a brief set of user inputs, it augments the file with additional entries, namely, \verb|z|, \texttt{A$_\texttt{V}$}, \verb|RA|, \verb|DEC|, \verb|minSNR| and \verb|target_SNR|, where \texttt{minSNR} and \verb|target_SNR| specify the minimum and target SNR for the Voronoi binning. Additionally, the user will be prompted to provide the M$_{\bullet}$ mass, which they may choose to leave blank (this value is only required if they intend to perform dynamical modelling). At this stage, foreground stars are masked, Voronoi binning of the sample is performed, and the user is prompted to complete the configuration file \verb|glance_config.ini|, specifying all input parameters (paths, stellar templates, and other options; see the documentation for the full list). Instructions on how to complete the configuration file will be printed to the terminal.

The script is executed as:

\begin{verbatim}
$ python GLANCE_run_1st_time.py gal_list
\end{verbatim}

where \verb|gal_list| refers to the name of the input text file (e.g. \verb|gal_info.txt|). 
\\

Subsequently, \glance\ can be executed by the command:

\begin{verbatim}
$ python run_GLANCE.py input_parameters
\end{verbatim}

where \verb|input_parameters| correspond to:

\begin{itemize}[noitemsep, topsep=0pt, label=$\cdot$]
  \item \verb|-n / --ngal| — index of the galaxy in the input catalogue (i.e., \verb|gal_list|).
  \item \verb|-f / --functL| — comma-separated list of functions/modules to be executed in this run.
\end{itemize}
\

The keyword list supplied to the option \verb|--functL| determines which modules of the pipeline are executed. Possible options are:

\begin{itemize}[noitemsep, topsep=0pt, label=$\cdot$]
  \item \verb|VorBin| – performs Voronoi binning of the IFU data according to the preferred SNR values.\
  \item \verb|pPXF| – extracts stellar and/or gaseous kinematics using \ppxf\  with MILES and IUS templates.\
  \item \verb|Starlight| – performs spectral synthesis with \starlight\ using BC03, CB19 and MILES stellar libraries.\
  \item \verb|FADO| – performs spectral synthesis with \fado\ using BC03 and CB19 libraries.
  \item \verb|RemoveYoung| – applies the \RY\ algorithm to generate stellar population maps cleaned of young stars for eight age thresholds.
  \item \verb|BayesLOSVD| – recovers stellar kinematics with \bayes.
  \item \verb|SINFONI| – processes SINFONI near-IR IFU data, including \ppxf\ kinematics with GNIRS and NIFS templates.
  \item \verb|prep-Dyn.pPXF| – prepares input files for \dyn\ using the \ppxf\ kinematic maps.
  \item \verb|Dyn.pPXF-a| – dynamical modelling with \dyn@\ppxf: first iteration.
  \item \verb|Dyn.pPXF-b| – dynamical modelling with \dyn@\ppxf: full grid exploration.
  \item \verb|Dyn.pPXF-c| – dynamical modelling with \dyn@\ppxf: final production run.
  \item \verb|prep-Dyn.Bayes| – prepares input files for \dyn\ based on \bayes\ kinematics.
  \item \verb|Dyn.Bayes-a| – dynamical modelling with \dyn@\bayes: first iteration.
  \item \verb|Dyn.Bayes-b| – dynamical modelling with \dyn@\bayes: full grid exploration.
  \item \verb|Dyn.Bayes-c| – dynamical modelling with \dyn@	\bayes: final production run.
  \item \verb|Phot.Decomp| – iterative photometric decomposition using \textsc{PhotUtils} and \textsc{GALFITools}.
  \item \verb|Orb.Decomp| – performs orbital decomposition of the best-fit \dyn\ model.
\end{itemize}

This structure ensures that each \glance\ module can be flexibly activated, skipped, or combined in any order, allowing the user to tailor the analysis workflow to their scientific goals.

\vspace{2em}

As an illustration, a typical execution of \glance\ may look as follows:
\begin{verbatim}
$ python run_GLANCE.py -n 3 \
  -f Phot.Decomp,VorBin,FADO,RemoveYoung,pPXF
\end{verbatim}

This command runs \glance\ for the fourth galaxy in the input file,  sequentially activating the \verb|Phot.Decomp|, \verb|VorBin|, \verb|FADO|, \verb|RemoveYoung|, and \verb|pPXF| modules. When executed, the pipeline creates a directory tree to organise the input and output of each module, for example \verb|NGC1566_analysis/FADO/BC03/IN|, \verb|.../OUT|, \verb|.../MC|, and analogous subfolders for the remaining modules. For a comprehensive overview of the available options and defaults, we strongly recommend the user to carefully inspect \glance's documentation. In addition, users are encouraged to review the scripts to better understand their workflow and, if necessary, customise it to their specific needs.

\newpage
\section{Showcasing the Analysis of \exG}\label{res}

This section presents the primary visual outputs of the pipeline for \exG\footnote{\textbf{NGC 1566}: Flocculent intermediate spiral galaxy at $\sim$31° inclination \citep{Ela19}. Dynamical modelling reveals a kinematically cold stellar component in the nuclear region, tracing a dynamically cool stellar disk. Stellar kinematic maps show a prominent nuclear spiral structure, consistent with patterns seen in some emission-line maps and velocity dispersion profiles. The galaxy lacks a strong large-scale bar, though several studies suggest a weak/intermediate-strength bar in the central region \citep[e.g.,][]{Com14,Sla19}. Residing in the Dorado Group (5+ galaxies), \exG\ hosts a Seyfert nucleus with no significant evidence of recent tidal interactions \citep{Sma15,Ela19}.}, illustrating its capability to analyse spectral and photometric datasets through an exemplary galaxy. While a much larger IFS mosaic of this galaxy exists, here we focus on a very high-quality dataset -- with a target SNR of 200 between 6390 and 6490 \AA\ -- which is sufficient to demonstrate the functionalities and outputs of the pipeline. Once the module for bar modelling is incorporated in future versions, we plan to re-process this dataset in order to assess the possible influence of \exG's weak bar on the derived orbital structures, which in the present analysis are obtained under the assumption of axisymmetry.

This section includes the results from $K$-band surface photometry, spectral synthesis, stellar kinematics extraction, and subsequent dynamical modelling. These plots serve as a visual guide for pipeline users and provide a clear expectation of the pipeline's outputs. 

\subsection{$K$-band Surface Photometry}

In this subsection, we present and describe the main output from the $K$-band surface photometry module for \exG, having as main goal to define the galaxy's central region. This delineation is crucial for comparing the properties of the galaxy's centre with those of its periphery in subsequent analyses.

\begin{figure*}
\centering
\includegraphics[trim={2cm 0 1.5cm 0}, clip, width=1\linewidth]{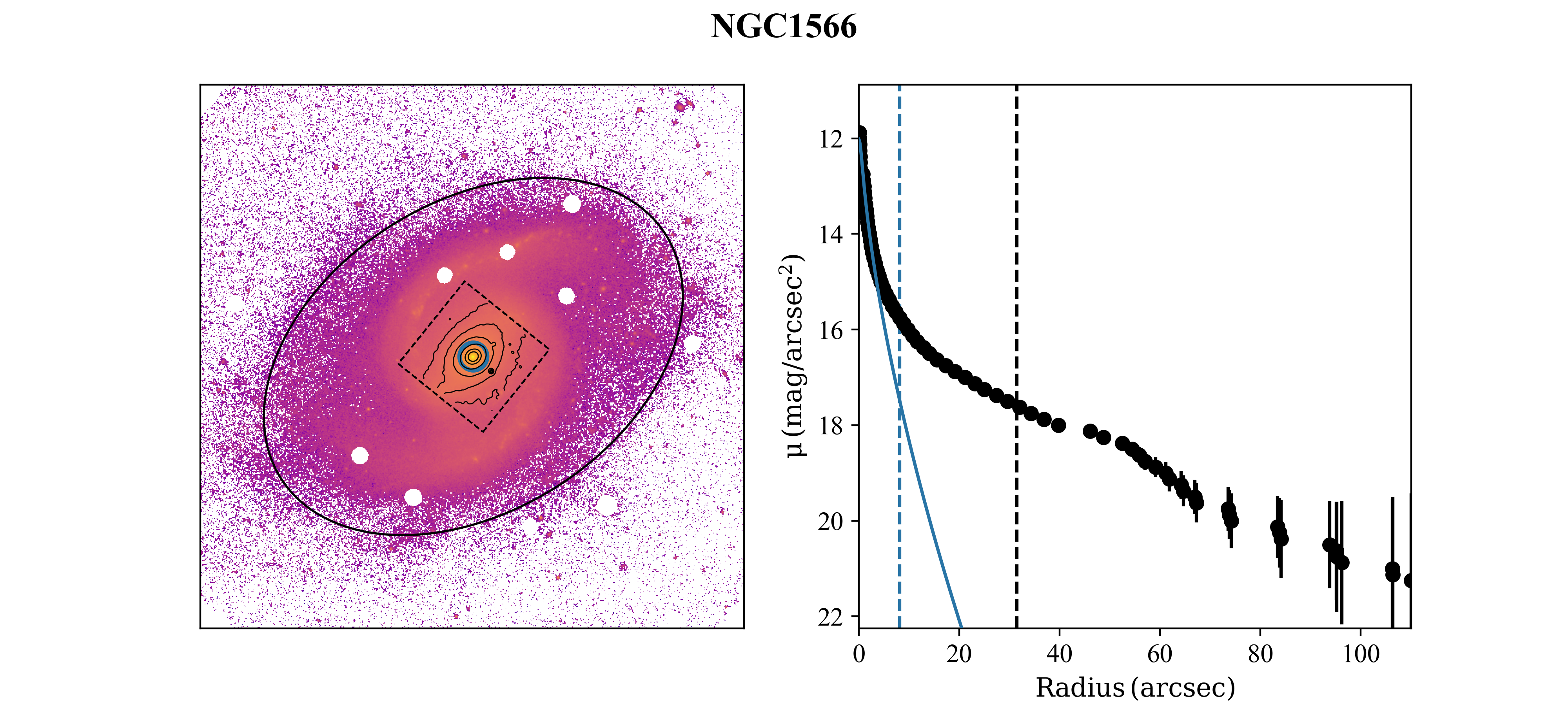}
\caption{Left panel: $K$-band photometric image of \exG, as observed by VISTA (VHS), rotated to the kinematic PA. The FoV of the MUSE observation is indicated by a dashed square, with contours highlighting the morphology within this region. 
The blue circle indicates the central region as assessed by the S\'ersic fit, while the ellipse illustrates the widest point in the SBP, outlining the galaxy's size. Right panel: Derived SBP with a blue line indicating the best-fitting S\'ersic model to the central luminosity excess. The blue dashed vertical line marks the radius derived from the Sérsic fit, while the black dashed line indicates MUSE's FoV.}
\label{sbp}
\end{figure*}

Figure~\ref{sbp} displays the main visual output of the surface photometry procedure. On the left-hand side, the sky-subtracted $K$-band photometric image of \exG\ observed by VISTA (VHS) is shown, with the MUSE's FoV indicated by a dashed square. The small and large circles mark the central and galaxy radii, respectively, with the latter being defined as the point where the SBP reaches the sky level, or where the $\sigma$ of the SBP surpasses 2 mag. On the right-hand side, the derived SBP is displayed, with the blue line indicating the best-fitting S\'ersic model to the central luminosity excess, accounting for PSF convolution effects. The vertical dashed line symbolizes the characteristic central radius determined by the S\'ersic fit.

\subsection{Stellar Population Synthesis}

\begin{figure*}
    \centering
    \includegraphics[width=0.9\linewidth]{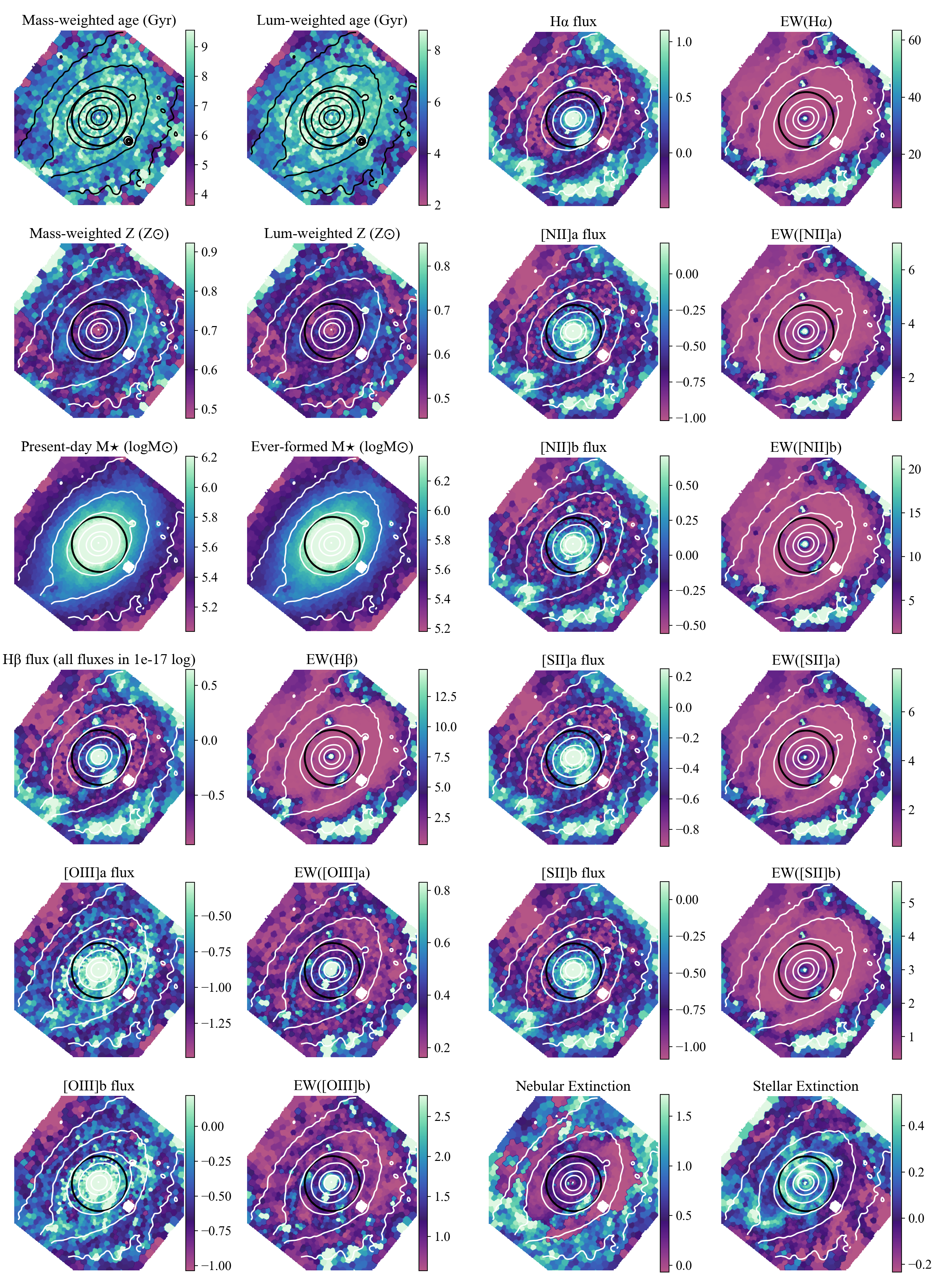}
    \caption{FADO Spectral Synthesis Results for \exG\ (BC03 SSP Library). The figure presents 24 maps, displaying: 1) Stellar population properties, including mass- and luminosity-weighted ages (Gyr) and metallicities (Z$_\odot$), present-day and total ever-formed stellar masses (logM$_\odot$), and stellar surface density; 2) Emission-line fluxes (with units of $10^{-17}$ $\cdot$ erg s$^{-1}$ cm$^{-2}$, in log-space) and equivalent widths (\AA) for H$\beta$, [OIII]$\lambda\lambda$4959,5007, H$\alpha$, [NII]$\lambda\lambda$6548,6583, and [SII]$\lambda\lambda$6717,6731; 3) Extinction maps, including $E(B-V)_{\text{neb}}$ and $A_{\rm V}$. Similar figures are produced for other SSP libraries (e.g., CB19) and for \starlight, though the latter excludes extinction maps. Photometric contours are overplotted, and the black circle represents the characteristic central region.}
    \label{fado0}
\end{figure*}

\begin{figure*}
    \centering
    \includegraphics[trim={2cm 0 2cm 0}, clip, width=\linewidth]{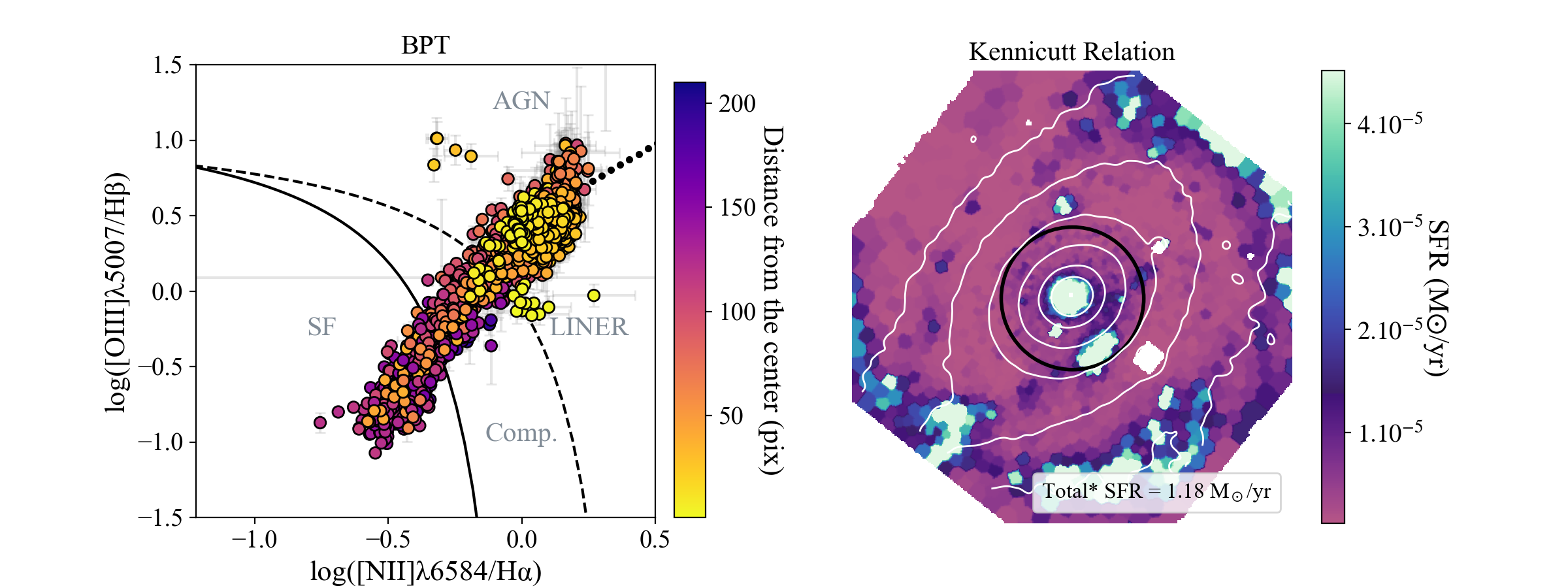}
    \caption{BPT diagram and Kennicutt relation for \exG. Left panel: BPT diagram with emission-line ratios [NII]$\lambda$6584/H$\alpha$ vs. [OIII]$\lambda$5007/H$\beta$, color-coded by distance from the galaxy centre. Demarcation lines from \citet{Kew01}, \citet{Sch07}, and \citet{Kau03} are overlaid to classify regions as star-forming, AGN-dominated, or composite. Right panel: spatially resolved star-formation rate (SFR) derived from the Kennicutt relation using H$\alpha$ luminosity, with values in units of M$_\odot$/yr. The central region is demarcated by a black circle and the map is rotated to the kinematic PA. The total estimated SFR is shown at the bottom of the figure (an asterisk (*) indicates that the MUSE FoV does not fully cover the stellar disk).}
    \label{bpt}
\end{figure*}

SPS via \fado\ and \starlight\ is performed, utilizing two distinct SSP libraries for \fado\ (BC03 \& CB19), and three for \starlight\ (BC03, CB19 \& MILES). The spectral synthesis results for \exG\ are summarized in Fig.~\ref{fado0}, which displays 24 maps derived from the \fado\ analysis using the BC03 SSP library. The top, left-side rows show the spatial distribution of stellar population properties, including mass- and luminosity-weighted ages and metallicities, as well as present-day and total ever-formed stellar masses. Bottom, left and right-side rows highlight emission-line fluxes and equivalent widths for key emission-lines. The right side of the bottom row provides extinction maps for both nebular and stellar components. Similar figures are generated for other SSP libraries and for \starlight, though the latter does not include extinction maps. If the user enables MC simulations, additional images (besides \texttt{fits} and \texttt{pickle} files) are produced, displaying the mean and standard deviation of the properties estimated by SPS. In addition, the BPT diagram and a map of the star-formation rate (SFR) via the Kennicutt relation are presented in Fig.~\ref{bpt}. The left panel shows the distribution of emission-line ratios [NII]$\lambda$6584/H$\alpha$ vs. [OIII]$\lambda$5007/H$\beta$, color-coded by distance from the galaxy center. Demarcation lines from \cite{Kew01}, \cite{Sch07}, and \cite{Kau03} are overlaid to classify regions as star-forming, AGN-dominated, or composite. The right panel displays the spatially resolved star formation rate (SFR) derived from the Kennicutt relation (M$_\odot$/yr). This figure provides insights into the ionization mechanisms and star formation activity across the central region of \exG.

Although a detailed comparison of spectral synthesis results using different stellar libraries remains to be conducted, inspection of the different maps indicate that the CB19 library consistently yields overestimated stellar ages for both \fado\ and \starlight. This systematic discrepancy may arise from the CB19 library's larger number of SSPs, which could introduce degeneracies in the fitting process. 

\subsection{Dynamical Analysis}

\subsubsection{Kinematic Extraction}

\begin{figure}
\centering
\includegraphics[trim={1cm 0 1cm 0}, clip, width=\linewidth]{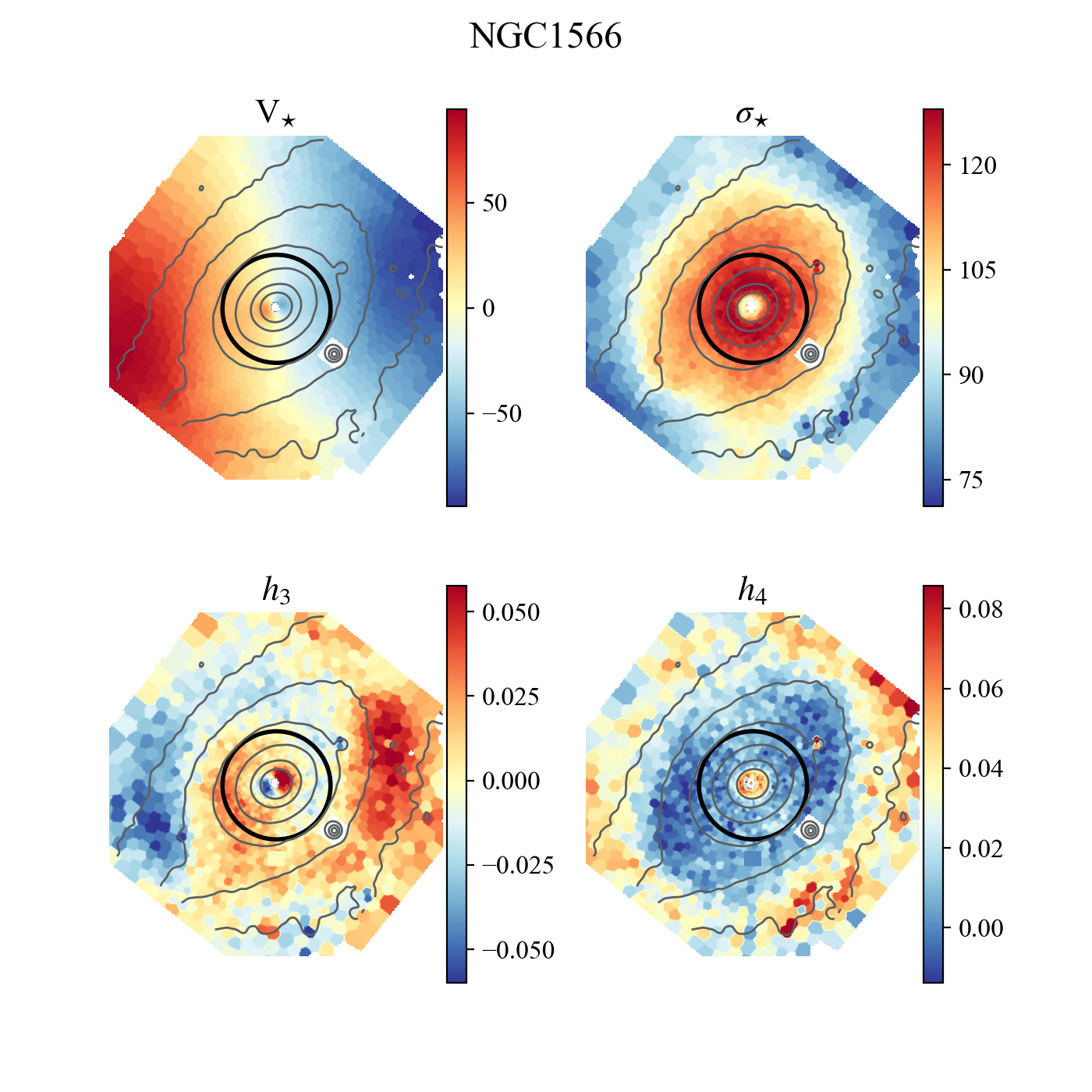}
\caption{Stellar kinematic maps for \exG, displaying the mean velocity, velocity dispersion, and higher-order moments (h$_3$ and h$_4$). Maps are rotated to the kinematic PA, with color bars in units of km/s. The black circle represents the central region as assessed by $K$-band surface photometry.}
\label{ppxf}
\end{figure}

\begin{figure}
\centering
\includegraphics[width=\linewidth]{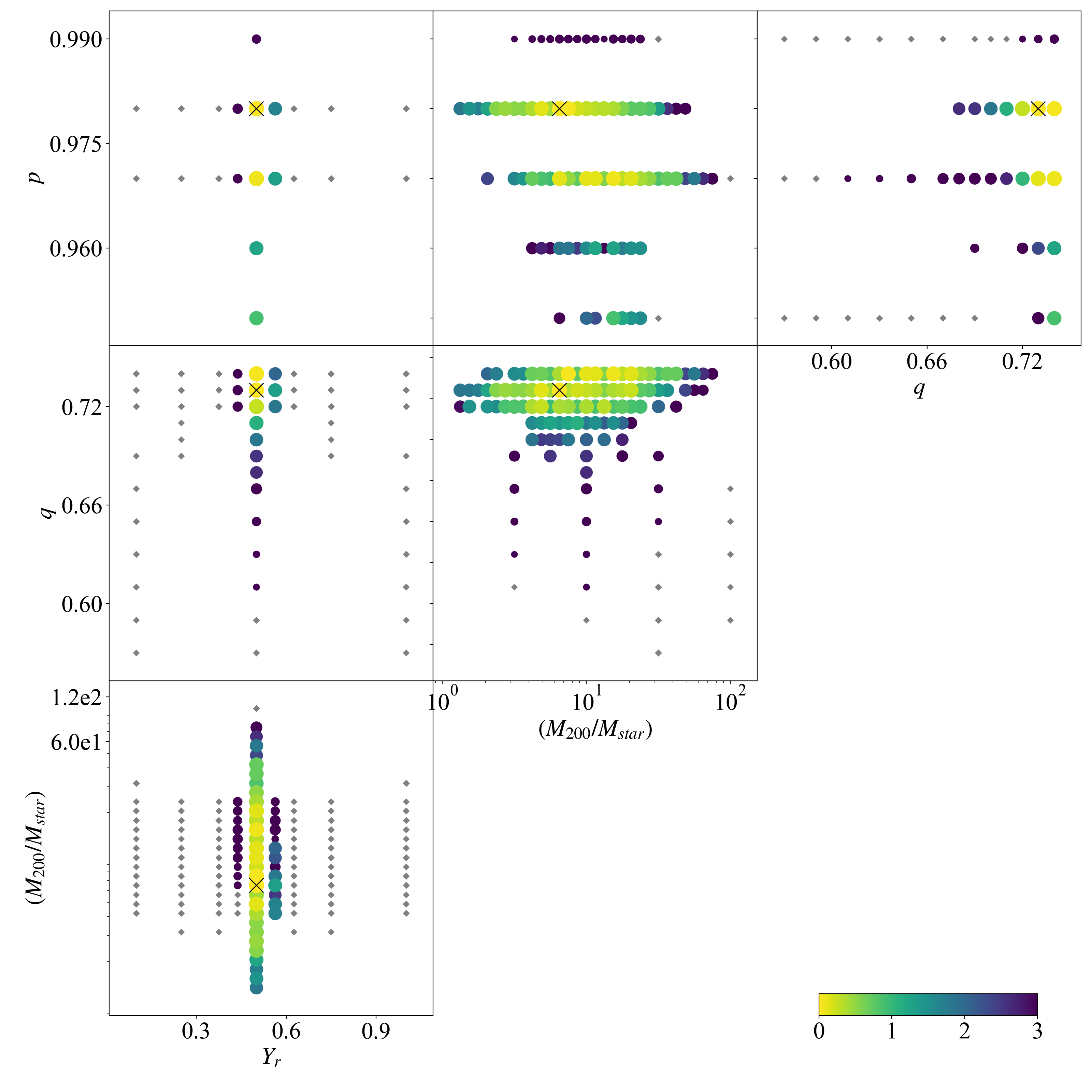}
\caption{Relationships between the free hyper-parameters ($p$, $q$, M$_{200}$, and \mlr) explored during the dynamical modelling process of \exG, demonstrating convergence of the dynamical model. The best-fit model is marked with a black cross. The coloured circles represent models within 3 sigma confidence level (light colours and larger circles indicate smaller values of the $\chi^2$). The small black dots indicate the models outside this confidence region. The best best-fitting values are $p$ = 0.98; $q$ = 0.73; \mlr\ = 0.5; M$_{200}$ = 6.49.}
\label{dyn1}
\end{figure}

Kinematic extraction with \ppxf\ is performed, adopting 2 different stellar libraries (IUS \& MILES), adopting the purely stellar setup, and the gas + stars, by combining the stellar and gaseous templates into a single array. In addition, the nebular continuum contribution is accessed and, in one case subtracted prior of fitting, while in the other is left unchanged. Upon inspection of the resulting stellar velocity maps, we verify that there are no significant differences between the various setups, selecting the IUS, neb.-cont. subtracted, gas + stellar fit for the dynamical modelling.


Figure~\ref{ppxf} presents the stellar kinematic maps of \exG\ derived from \ppxf\ analysis, showing the four primary velocity moments of the LOSVD. A compact nuclear disk is clearly detected in the galaxy’s inner region, as evidenced by the coherent structure in both the mean velocity and h$_{3}$ maps, in agreement with previous findings \citep{SaFre23}. Notably, the h$_{3}$ map exhibits a spiral pattern, likely tracing non-circular motions associated with the spiral arms. Furthermore, the tilted distribution of low h$_{4}$ values suggests the presence of an additional kinematic component, which may also be linked to the weak bar or spiral structure.




\subsubsection{Dynamical Modelling}

In subsequent steps, dynamical modelling by means of \dyn\ is carried out. The modelling process tries to optimize the galaxy's intrinsic shape (i.e., the intermediate‐to‐major axis ratio, $p$ = $b$/$a$, the minor‐to‐major axis ratio, $q$ = $c$/$a$, and the major axis's intrinsic-to-apparent ratio $u$ = $a'$/$a$) and gravitational potential (including halo mass, M$_{200}$, M$_{\bullet}$, and \mlr), by minimizing $\chi^2$ statistics. In the case of \exG, $u$ and log$_{10}$(M$_{\bullet}$/M$_{\rm sun}$) were fixed to $\approx$ 1 (as expected for a face-on, spiral galaxy) and 7.11 (\citealt{Mbh}), respectively. Initially, \dyn\ was executed only with the light MGE, implying a constant mass-to-light ratio \mlr\ across the galaxy, and subsequently with both light and mass MGEs, allowing for a spatially varying \mlr. As evidenced by Fig. \ref{dynSBPs}, the two setups yield consistent results.

\begin{figure*}
\centering
\includegraphics[width=\linewidth]{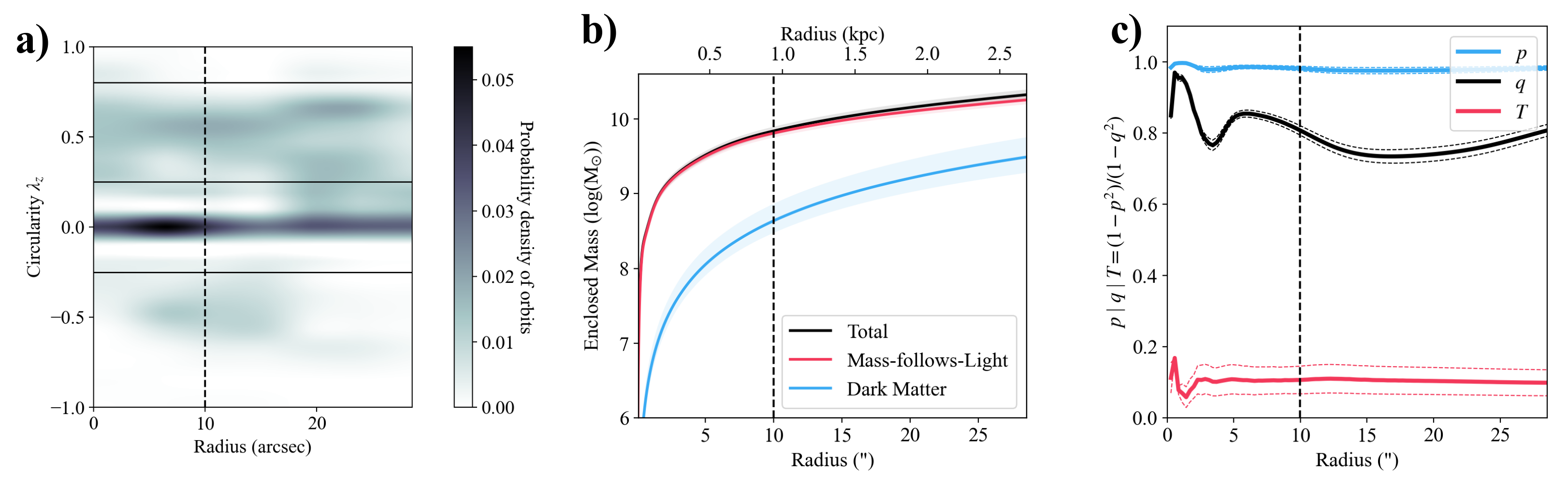}
\caption{Panel a) circularity plot showing the distribution of stellar orbits in the best-fit dynamical model of \exG. Horizontal lines denote the separations between cold, warm, hot, and counter-rotating orbits, with default cut values of $\lambda_{z} > 0.8\ $; $0.25\ < \lambda_{z} < 0.8\ $; $-0.25\ < \lambda_{z} < 0.25\ $; $\lambda_{z} < -0.25\ $, respectively. Panel b) enclosed mass profiles for the stellar (mass-follows-light) component (red), dark matter (blue), and their sum (black). Solid lines indicate best-fit models and shaded regions denote 1$\sigma$ uncertainties. Panel c) intrinsic axis ratios $q$ (blue) and $p$ (black) are shown as functions of distance from the galactic center. The triaxiality parameter T is plotted in red, quantifying the galaxy's 3D shape. Vertical dashed lines indicate the characteristic central radius.}
\label{dyn2}
\end{figure*}

It follows a brief description of the main visual outputs generated by \dyn\footnote{\dyn\ produces a variety of diagnostic plots beyond those presented here. For a comprehensive overview of its capabilities and outputs, we refer the reader to the original article.}: Fig.~\ref{dyn0} displays the observed kinematic moments, the best-fit model moments, and respective residuals. Additionally, relationships between the free parameters are presented in Fig.~\ref{dyn1}, providing insights into the optimization process.
Finally, panel a) of Fig.~\ref{dyn2} shows the circularity plot ($\lambda_{z}$), offering a detailed view of the system's dynamical structure, panel b) displays the radial distribution of the enclosed mass, and panel c) offers a perspective on the system's geometry, with $q$ and $p$ corresponding to the intrinsic flattening ($c$/$a$ and $b$/$a$, respectively), and T to the triaxiality of the system; a T = 1 ($a$ > $b$ = $c$; $q$ = $p$) corresponds to a prolate shape, while T = 0 ($a$ = $b$ > $c$; $p$ = 1) corresponds to an oblate shape. Inspection of these figures suggests that:

\begin{figure*}
\centering
\includegraphics[width=\linewidth]{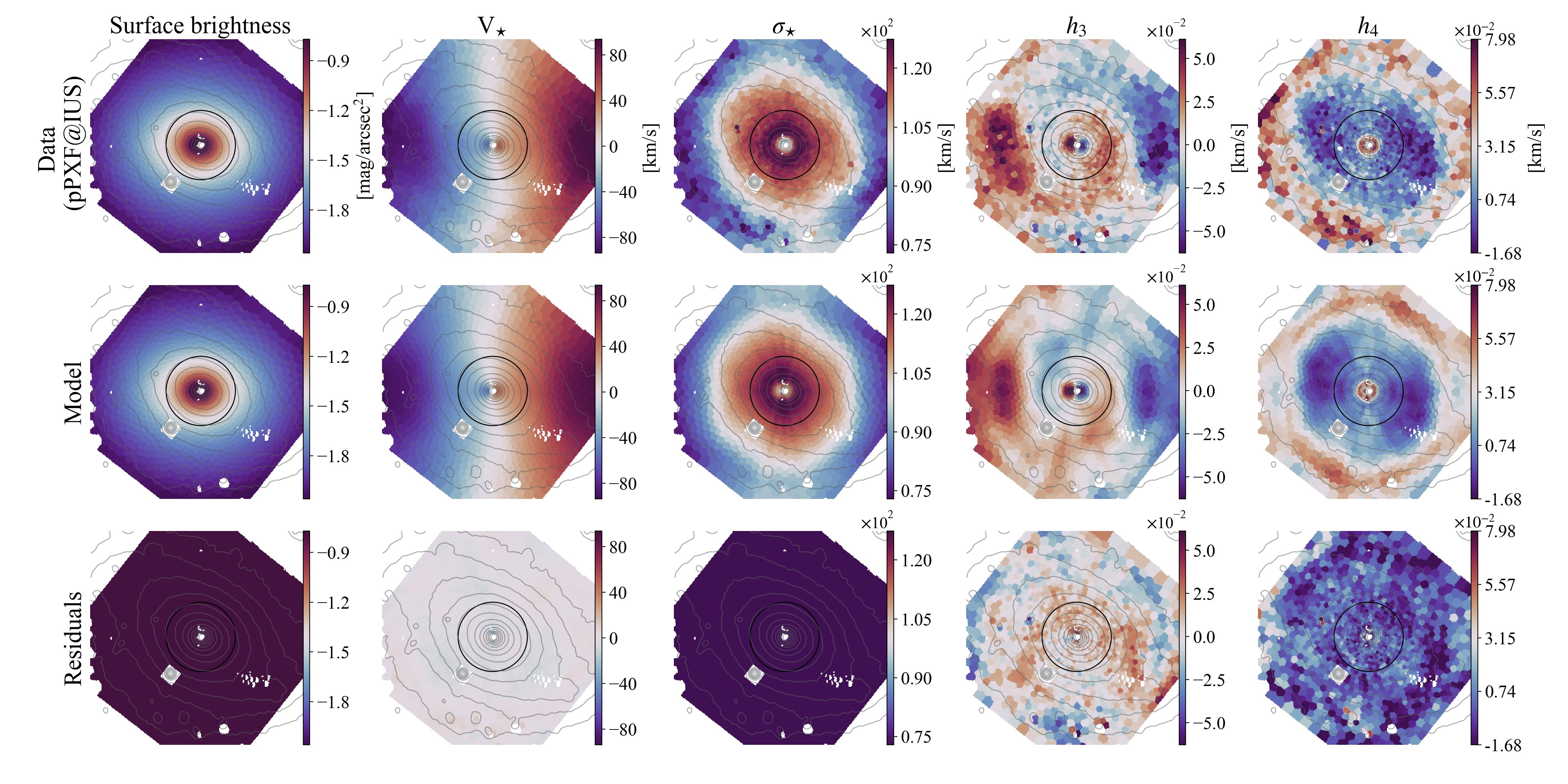}
\caption{Comparison of observed velocity moments (top row), best-fit model moments (middle row), and residuals (bottom row) from the \dyn\ best-fit Schwarzschild model for \exG. The residuals are normalized to their respective uncertainties (except for the surface brightness, which is normalized to the observed) to highlight deviations relative to the observational errors. Photometric contours are overplotted and the central region is demarcated by a black circle.}
\label{dyn0}
\end{figure*}

\vspace{-0.1em}

$\bullet$ Within the characteristic central radius, the density is dominated by hot and warm orbits -- signatures of a kinematically hot, classical bulge. However, within the same region, a non-negligible density at $\lambda_{z} \sim$ 0.9 is evident, hinting at the presence of an embedded cold component coexisting within the hot bulge (inspection of Fig.~\ref{dynSBPs} reveals an upturn of the cold component within this region $\sim$ 1.5 mag fainter than the total). Beyond this radius, the cold band strengthens to form the inner section of the main stellar disk. The fact that the hot component permeates the whole sampled region might be attributed to the fact that the MUSE observation primarily samples the bulge region, or it might reflect the stellar halo \citep{Zhu20}. The presence of counter-rotating orbits might suggest past merger events \citep{San24}, or originate from the weak bar \citep[e.g.,][]{Zei01,Zot18}.\\
$\bullet$ Within the region sampled by the MUSE observations, \exG\ is baryon-dominated.\\
$\bullet$ The distribution of $p$ and $q$ reveal that intermediate and major axes are nearly equal, evidencing a quasi axisymmetric morphology within the inner disk plane, and a moderately oblate central stellar component.\




\subsection{Orbital Decomposition}\label{res1}

Orbital decomposition is performed based on the distribution of the circularity parameter, enabling to distinguish the various dynamical components of the galaxy. The orbital decomposition module reconstructs maps of the cold, warm, and hot stellar components, which are then converted into radial profiles, offering a physically motivated representation of the radial luminosity profiles of the different dynamical components within the galaxy. 

\begin{figure*}
\centering
\includegraphics[width=1\linewidth]{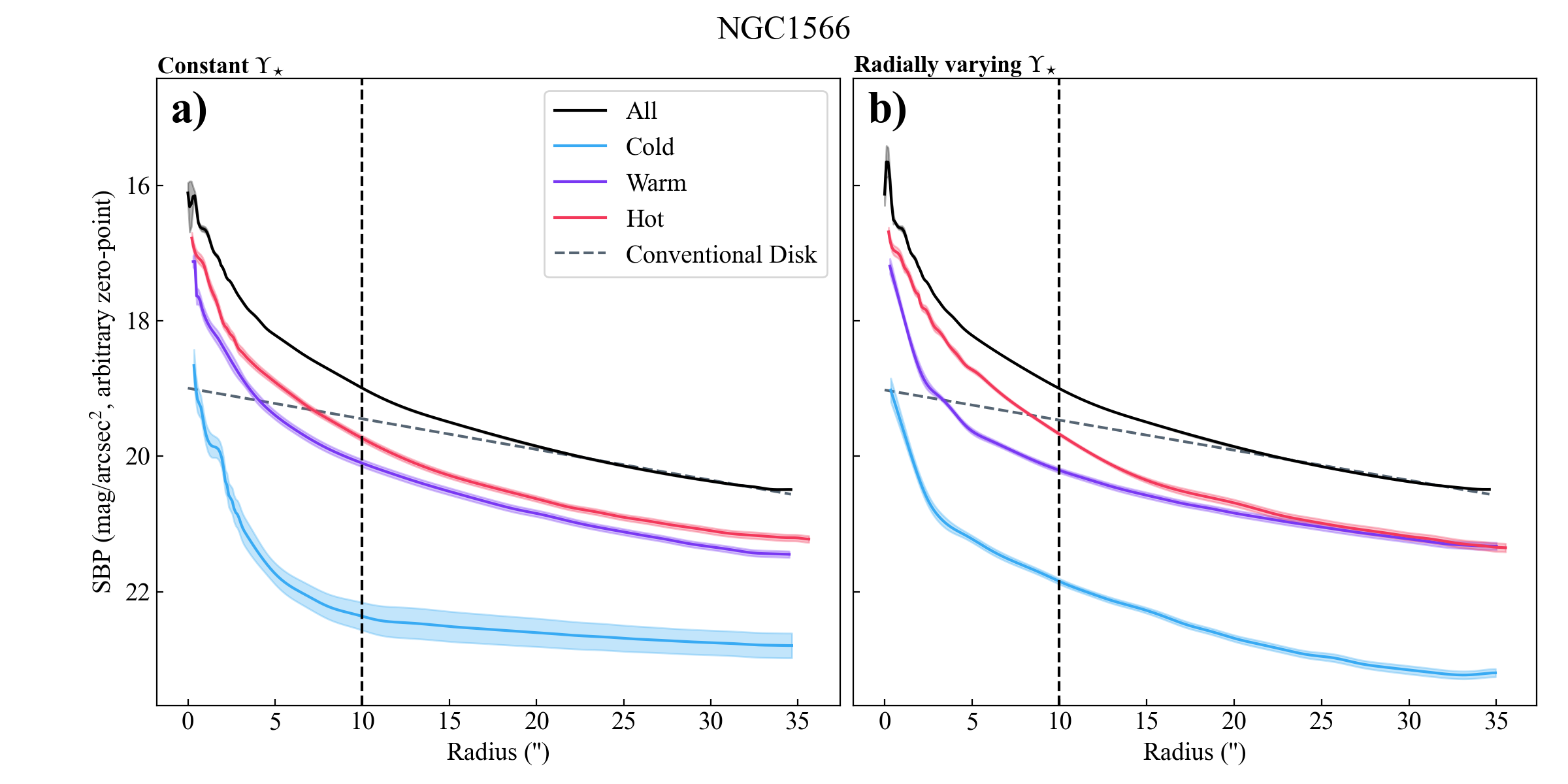}
\caption{SBPs of the different dynamical components of \exG, obtained through dynamical modelling and subsequent decomposition. The shaded regions indicate 1$\sigma$ uncertainties, derived from models whose $\chi^2$ values lie within 1$\sigma$ of the best-fit solution. Panel a) displays the SBPs of the cold (blue), warm (purple) and hot (red) stellar components as obtained with a constant \mlr, while panel b) displays the retrieved radial profiles adopting a radially varying \mlr. The grey line represents the conventional exponential disk model, derived from fitting the (total) SBP outside the bulge-dominated region. The vertical dashed line depicts the characteristic central boundary as obtained trough $K$-band surface photometry.}
\label{dynSBPs}
\end{figure*}

Figure~\ref{dynSBPs} displays the SBPs determined by decomposing the best-fit combination of orbits trough dynamical modelling for \exG, with the black, red, purple and blue lines representing the all, hot, warm, and cold orbital structures, respectively. Panel a) presents the results assuming a constant \mlr, while panel b) corresponds to a radially varying \mlr. In addition, the conventional exponential disk fit is shown as a dashed grey line. As in previous cases, the vertical line corresponds to the characteristic central radius, as defined by $K$-band surface photometry.

Inspection of this figure reveals an increase in this galaxy's cold orbit distribution towards the galactic centre due to the presence of a compact disk already appreciable from the stellar kinematic maps. Comparison of the recovered cold-disk profile (blue) with the traditional exponential model (grey) highlights the inaccuracy of assuming such a simplistic form: while this practice would overestimate the disk's contribution outside the bulge region, it would underestimate it within the bulge. However, caution is warranted, as the nuclear disk appears to constitute an additional, distinct cold component, rather than being part of the outer parent disk, concealing the true structure of the central parent cold-disk. 
Moreover, if one defines the disk as the sum of the cold and warm components, the exponential assumption remains inadequate, as the combined profile is supra-exponential, more closely resembling a S\'ersic profile with an index $\eta$ $\sim$ 2. 
However, to define the disk as the sum of the cold and warm components is by itself ambiguous, considering that it remains to be demonstrated if all orbits with 0.25 < $\lambda_{\rm z}$ < 0.8 belong in fact to the thick disk. 
This result challenges the applicability of a universal exponential disk model, as the assumption of an exponential profile does not accurately describe the observed distributions. 

\section{Summary and conclusions}\label{conc}

In this work, we present a comprehensive and publicly available pipeline designed to perform galactic archaeology, aimed at constraining galaxy formation and evolution scenarios by combining spatially resolved star formation histories, stellar dynamics, emission-line properties, and surface photometry, within a single framework. Although our primary motivation for its development is to reassess the conventional assumption that the exponential nature of the disk extends into the inner regions of spiral galaxies, this pipeline can be used to explore numerous science cases, such as, for instance, the mass assembly histories of galaxies, the connection between stellar populations and internal dynamics, the structural decomposition of bulges and disks, and the role of central components such as bars and supermassive black holes in shaping galaxy evolution. This pipeline was applied to a set of eight galaxies with exceptional-quality MUSE data, offering a comprehensive investigation of their stellar structures (Breda et al., submitted for publication in A\&A).


The developed pipeline integrates surface photometry to define a characteristic central radius, SPS of MUSE IFS data to derive SFHs, kinematic extraction, and dynamical modelling. By automating the application of various analysis techniques -- such as spectral synthesis (\fado, \starlight), post-processing with \ry, kinematic extraction (\ppxf, \bayes), and dynamical modelling (\dyn) -- it offers a robust framework for the systematic study of galaxies. Additionally, it streamlines essential pre-processing steps, including Galactic extinction correction, de-redshifting, Voronoi binning, and nebular continuum correction, while incorporating parallel processing to enhance computational efficiency. 

This work presents the first step towards a more complete framework for galactic archaeology, by developing a set of tools that provides a comprehensive view of galactic substructure, generating vast amounts of estimates across a wide range of physical properties for various sets of techniques, libraries and setups.
Looking ahead, future versions of the pipeline will incorporate additional modules featuring alternative methods to extract valuable information from IFS data, and to handle the orbit colouring functionality in \dyn\ for intuitive classification and analysis of orbital structures, as well as dedicated routines to prepare and process data for dynamical modelling including a bar component, ensuring full compatibility with the latest capabilities of \dyn. 
Furthermore, to fully exploit the richness of these outputs, we envision the development of an additional module that leverages AI-driven frameworks to uncover physically meaningful correlations and/or clustering patterns within \glance's outputs, considering that the complexity and dimensionality of the data make manual exploration impractical. For instance, in a recent application of \glance\ to NGC 4030 \citep{Bre24}, we identified a spatial correlation between stellar age and velocity dispersion, indicating that stars appear to form with low $\sigma_\star$ and are gradually heated over time by dynamical processes. Although this relation was identified without the use of AI, such physically meaningful trends may be more systematically and efficiently uncovered through such approaches. The development of an AI module will therefore greatly enhance the ability to detect and interpret such connections, providing new insights into the underlying mechanisms driving galaxy evolution. 

\section*{Acknowledgements}

I.B. has received funding from the European Union's Horizon 2020 research and innovation programme under the Marie Sklodowska-Curie Grant agreement ID n.º 101059532. This project was extended for 6 months by the Franziska Seidl Funding Program of the University of Vienna.
J.F-B acknowledges support from the PID2022-140869NB-I00 grant from the Spanish Ministry of Science and Innovation.
M.O. acknowledges support from JSPS KAKENHI, Grant Number: JP25K07361.
AFK acknowledges funding from the Austrian Science Fund (FWF) [grant DOI 10.55776/ESP542].
We thank  Dr. Dimitri Gadotti for his insightful comments and suggestions.
This research has made use of the NASA/IPAC Extragalactic Database (NED) which is operated by the Jet Propulsion Laboratory, California Institute of Technology, under contract with the National Aeronautics and Space Administration.

\section*{Data Availability}

The MUSE integral-field spectroscopic data used in this work are publicly available as part of the PHANGS-MUSE survey Data Release 1 \citep{phangs} and can be accessed via the link \url{https://www.canfar.net/storage/vault/list/phangs/RELEASES/PHANGS-MUSE/DR1.0}. The $K$-band imaging from the VISTA Hemisphere Survey \citep[VHS][]{vhs} is publicly available through the VISTA Science Archive facility at \url{hhttp://vsa.roe.ac.uk/dboverview.html}. The HST imaging in the F814W filter utilized to extract the light MGE was retrieved from the Mikulski Archive for Space Telescopes (MAST) at \url{https://mast.stsci.edu}. 



\bibliographystyle{mnras}

\input{References01.tex}



\appendix
\onecolumn
\section*{Appendix: Dynamite's diagnostic maps for \exG\ assuming a radially varying \mlr}

\centering
\includegraphics[width=\linewidth]{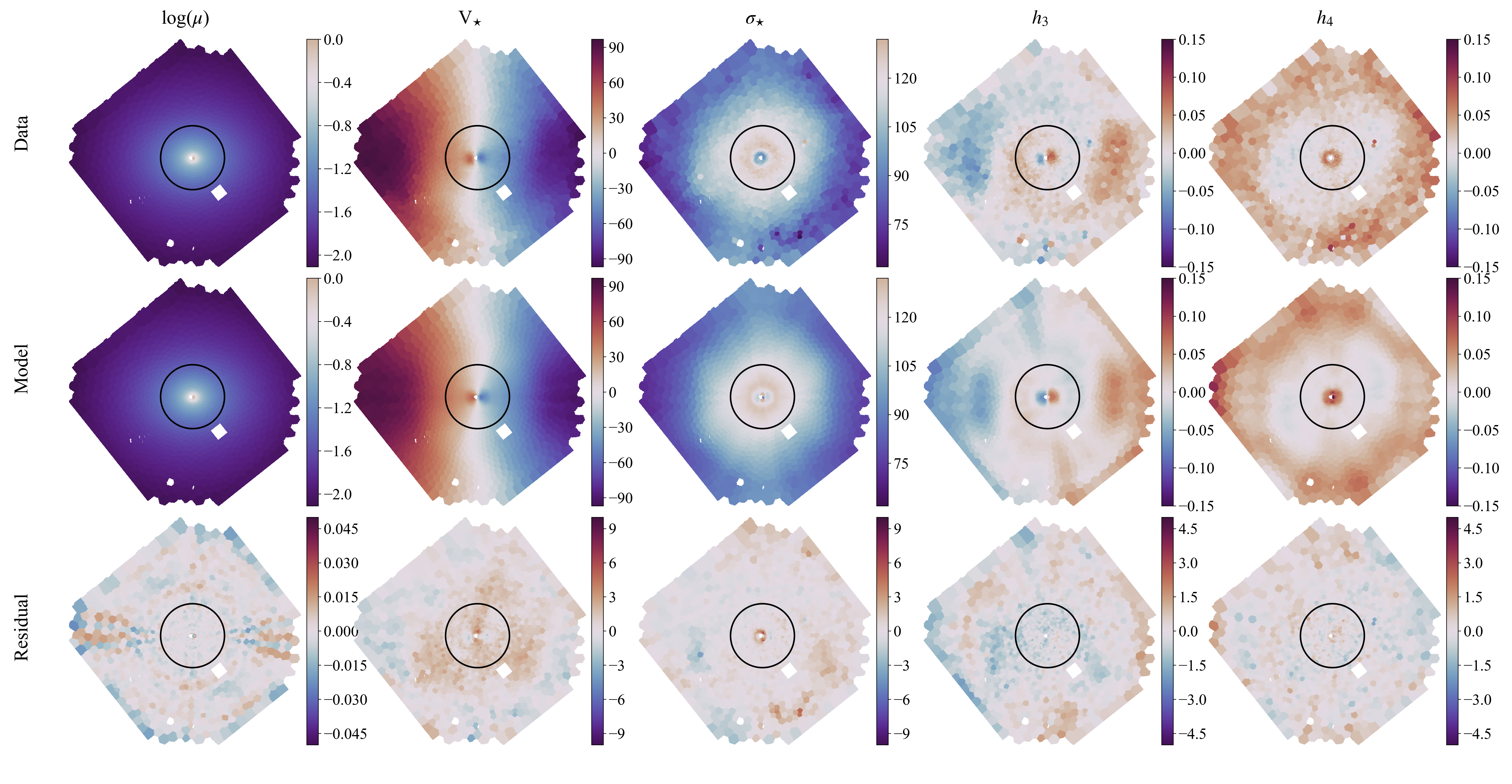}

\noindent \textbf{Figure A1.} Similar to Fig.~\ref{dyn0}, featuring a fit with radially varying \mlr.

\includegraphics[width=\linewidth]{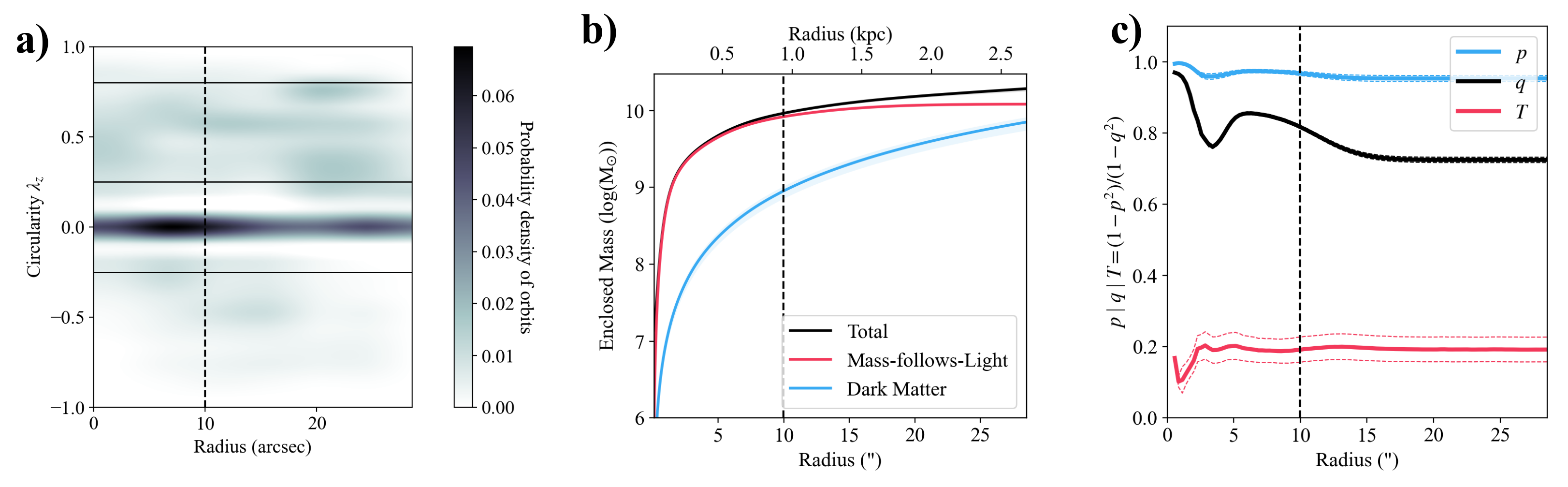}

\noindent \textbf{Figure A2.} Similar to Fig.~\ref{dyn2}, featuring a fit with radially varying \mlr.

\twocolumn

\begin{figure}
\centering
\includegraphics[width=\linewidth]{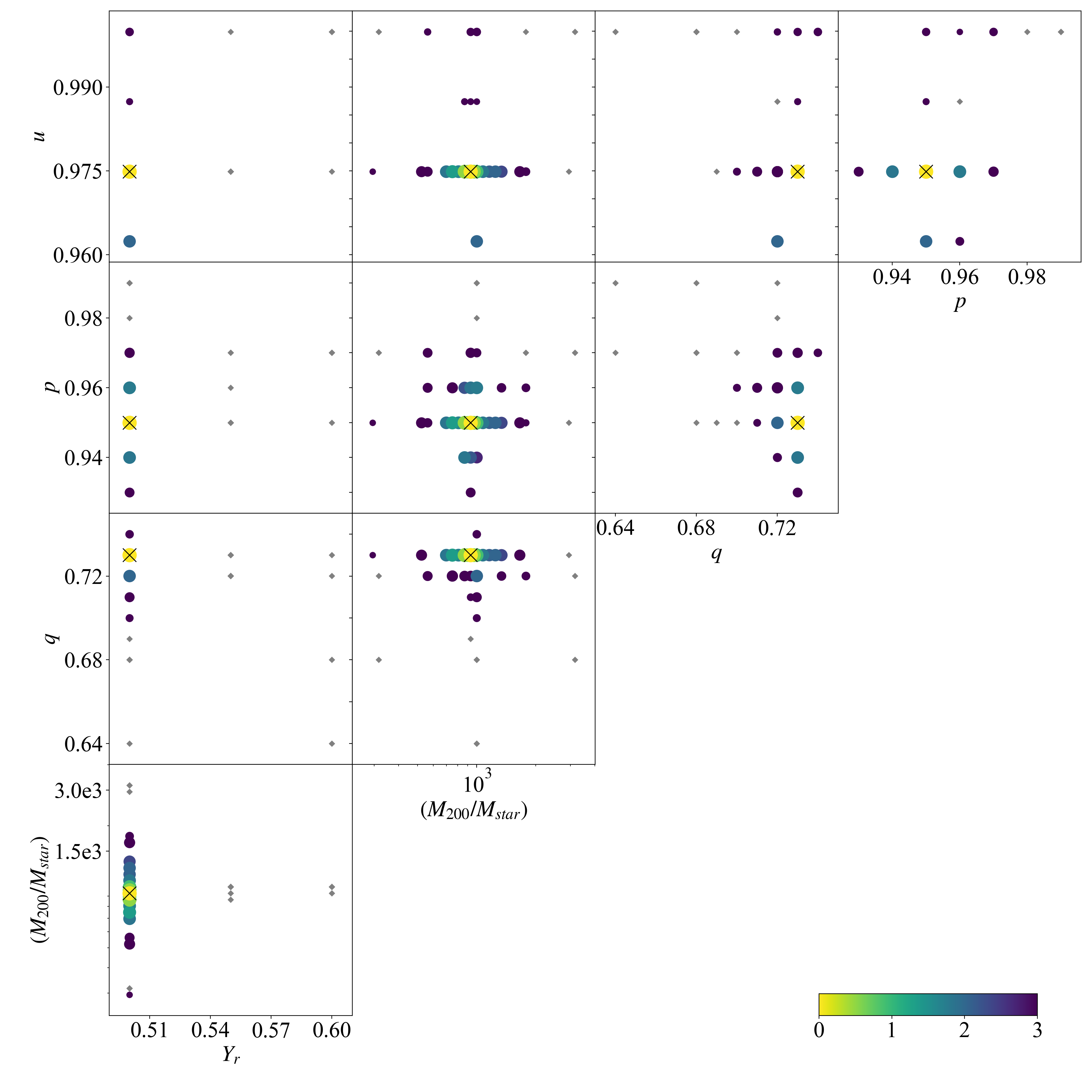}
\caption{Similar to Fig.~\ref{dyn1}, featuring a fit with radially varying \mlr.}
\label{dyn1A}
\end{figure}


\bsp	
\label{lastpage}
\end{document}

%% file: References01.tex

%% file: GLANCE_MNRAS_final.bbl
\begin{thebibliography}{}

\bibitem[Bacon et al.(2014)]{muse} Bacon R., Vernet J., Borisova E., Bouch{\'e} N., Brinchmann J., Carollo M., Carton D., et al., 2014, Msngr, 157, 13
\bibitem[Baldwin, Phillips \& Terlevich(1981)]{BPT} Baldwin, J. A., Phillips, M. M. \& Terlevich, R., 1981, PASP, 93, 5
\bibitem[Barnes \& Hernquist(1996)]{BarHer96} Barnes J. E., Hernquist L., 1996, ApJ, 471, 115
\bibitem[Bender \& M\"ollenhoff(1987)]{fitell3} Bender, R. \& M\"ollenhoff. C., 1987, A\&A, 177,71
\bibitem[Breda et al.(2024)]{Bre24} Breda I., van de Ven G., Thater S., Falc{\'o}n-Barroso J., Jethwa P., Gadotti D.~A., Onodera M., et al., 2024, A\&A, 692, L10. doi:10.1051/0004-6361/202452595
\bibitem[Davis et al.(2014)]{Mbh} Davis B.~L., Berrier J.~C., Johns L., Shields D.~W., Hartley M.~T., Kennefick D., Kennefick J., et al., 2014, ApJ, 789, 124. doi:10.1088/0004-637X/789/2/124
\bibitem[Bittner et al.(2019)]{gist} Bittner A., Falc{\'o}n-Barroso J., Nedelchev B., Dorta A., Gadotti D.~A., Sarzi M., Molaeinezhad A., et al., 2019, A\&A, 628, A117. doi:10.1051/0004-6361/201935829
\bibitem[Bournaud, Jog \& Combes(2005)]{BouJogCom05}Bournaud F., Jog C. J., Combes F., 2005, A\&A, 437, 69
\bibitem[Breda \& Papaderos(2018)]{BP18}Breda, I. \& Papaderos, P. 2018, A\&A, 614, 48
\bibitem[Breda et al.(2019)]{ifit}Breda, I., Papaderos, P., Gomes, J.M., Amarantidis, S. 2019, A\&A, 632, A128
\bibitem[Breda et al.(2020a)]{Bre20a} Breda I., Papaderos, P., Gomes, J.M., V\'ilchez, J.M., Ziegler, B.L., et al., 2020a, A\&A, 635, A177
\bibitem[Breda et al.(2020b)]{Bre20b} Breda I., Papaderos P., Gomes J. M., 2020, A\&A, 640, A20
\bibitem[Breda \& Papaderos(2023)]{BP23} Breda I., Papaderos P., 2023, A\&A, 669, A70. doi:10.1051/0004-6361/202245095
\bibitem[Breddels \& Helmi(2014)]{BreHel14} Breddels, M. A., \& Helmi, A. 2014, ApJL, 791, L3, doi: 10.1088/2041-8205/791/1/L3
\bibitem[Bressan et al.(2012)]{Bre12} Bressan, A., et al. 2012, MNRAS, 427
\bibitem[Bryant et al.(2015)]{sami} Bryant J.~J., Owers M.~S., Robotham A.~S.~G., Croom S.~M., Driver S.~P., Drinkwater M.~J., Lorente N.~P.~F., et al., 2015, MNRAS, 447, 2857. doi:10.1093/mnras/stu2635
\bibitem[Bruzual \& Charlot(2003)]{BruCha03} Bruzual, G. \& Charlot, S., 2003, MNRAS, 344, 1000
\bibitem[Bundy et al.(2015)]{manga} Bundy K., Bershady M.~A., Law D.~R., Yan R., Drory N., MacDonald N., Wake D.~A., et al., 2015, ApJ, 798, 7. doi:10.1088/0004-637X/798/1/7
\bibitem[Calzetti et al.(2000)]{Cal00} Calzetti D., Armus L., Bohlin R.~C., Kinney A.~L., Koornneef J., Storchi-Bergmann T., 2000, ApJ, 533, 682
\bibitem[Cappellari(2002)]{Cap02} Cappellari M., 2002, MNRAS, 333, 400
\bibitem[Cappellari \& Copin(2003)]{voronoi}  Cappellari M., Copin Y., 2003, MNRAS, 342, 345. doi:10.1046/j.1365-8711.2003.06541.x
\bibitem[Cappellari \& Emsellem(2004)]{CapEms04} Cappellari M., Emsellem E., 2004, PASP, 116, 138
\bibitem[Cappellari et al.(2006)]{Cap06} Cappellari M., Bacon R., Bureau M., Damen M.~C., Davies R.~L., de Zeeuw P.~T., Emsellem E., et al., 2006, MNRAS, 366, 1126. doi:10.1111/j.1365-2966.2005.09981.x
\bibitem[Cappellari(2017)]{Cap17} Cappellari M., 2017, MNRAS, 466, 798
\bibitem[Cid Fernandes et al.(2005)]{starlight} Cid Fernandes, R., Mateus, A., Sodr\'{e}, L., Stasi\'{n}ska, G., Gomes, J. M., 2005, MNRAS, 358, 363
\bibitem[Chen et al.(2015)]{Chen15} Chen, Y., et al. 2015, MNRAS, 452, 1068
\bibitem[Combes et al.(2014)]{Com14} Combes F., Garc{\'\i}a-Burillo S., Casasola V., Hunt L.~K., Krips M., Baker A.~J., Boone F., et al., 2014, A\&A, 565, A97. doi:10.1051/0004-6361/201423433

\bibitem[den Brok et al.(2021)]{dBro21} den Brok M., Krajnovi{\'c} D., Emsellem E., Brinchmann J., Maseda M., 2021, MNRAS, 508, 4786. doi:10.1093/mnras/stab2852
\bibitem[Dye et al.(2018)]{uhs} Dye S., Lawrence A., Read M.~A., Fan X., Kerr T., Varricatt W., Furnell K.~E., et al., 2018, MNRAS, 473, 5113. doi:10.1093/mnras/stx2622
\bibitem[Edge et al.(2013)]{viking} Edge A., Sutherland W., Kuijken K., Driver S., McMahon R., Eales S., Emerson J.~P., 2013, Msngr, 154, 32
\bibitem[Elagali et al.(2019)]{Ela19} Elagali A., Staveley-Smith L., Rhee J., Wong O.~I., Bosma A., Westmeier T., Koribalski B.~S., et al., 2019, MNRAS, 487, 2797. doi:10.1093/mnras/stz1448
\bibitem[Emsellem et al.(2022)]{phangs} Emsellem E., Schinnerer E., Santoro F., Belfiore F., Pessa I., McElroy R., Blanc G.~A., et al., 2022, A\&A, 659, A191. doi:10.1051/0004-6361/202141727
\bibitem[Falc{\'o}n-Barroso et al.(2006)]{gandalf} Falc{\'o}n-Barroso J., Bacon R., Bureau M., Cappellari M., Davies R.~L., de Zeeuw P.~T., Emsellem E., et al., 2006, MNRAS, 369, 529. doi:10.1111/j.1365-2966.2006.10261.x
\bibitem[Falc{\'o}n-Barroso et al.(2011)]{MILES} Falc{\'o}n-Barroso J., S{\'a}nchez-Bl{\'a}zquez P., Vazdekis A., Ricciardelli E., Cardiel N., Cenarro A.~J., Gorgas J., et al., 2011, A\&A, 532, A95. doi:10.1051/0004-6361/201116842
\bibitem[Fahrion et al.(2019)]{Fah19} Fahrion, K., Lyubenova, M., van de Ven, G., et al. 2019, A\&A, 628, A92, doi: 10.1051/0004-6361/201935832
\bibitem[Falc{\'o}n-Barroso \& Martig(2021)]{FalMar21} Falc{\'o}n-Barroso J., Martig M., 2021, A\&A, 646, A31
\bibitem[Feldmeier-Krause et al.(2017)]{Fel17} Feldmeier-Krause, A., Zhu, L., Neumayer, N., et al. 2017, MNRAS, 466, 4040, doi: 10.1093/mnras/stw3377
\bibitem[Fern\'andez\ Lorenzo et al.(2014)]{FerLor14}Fernandez\ Lorenzo, M., Sulentic, J., Verdes-Montenegro, L., et al. 2014, ApJL, 788, L39
\bibitem[Fisher \& Drory(2011)]{FisDro11} Fisher, D. \& Drory, N., 2011, ApJL, 733, L47
\bibitem[Fioc \& Rocca-Volmerange(1999)]{pegase} Fioc M., Rocca-Volmerange B., 1999, arXiv, astro-ph/9912179
\bibitem[Gadotti(2009)]{Gad09} Gadotti, D. A., 2009, MNRAS, 393, 1531
\bibitem[Ginsburg \& Mirocha(2011)]{pyspeckit1} Ginsburg A., Mirocha J., 2011, ascl.soft. ascl:1109.001
\bibitem[Ginsburg et al.(2022)]{pyspeckit2} Ginsburg A., Sokolov V., de Val-Borro M., Rosolowsky E., Pineda J.~E., Sip{\H{o}}cz B.~M., Henshaw J.~D., 2022, AJ, 163, 291
\bibitem[Gomes \& Papaderos(2016)]{RY}Gomes, J. M. \& Papaderos, P., 2016, A\&A, 594, A49
\bibitem[Gomes \& Papaderos(2017)]{fado}Gomes, J. M. \& Papaderos, P., 2017, A\&A, 603, A63, ESA SP-402. European Space Agency , Noordwijk , p. 621
\bibitem[Gordon et al.(2023)]{G23} Gordon K.~D., Clayton G.~C., Decleir M., Fitzpatrick E.~L., Massa D., Misselt K.~A., Tollerud E.~J., 2023, ApJ, 950, 86. doi:10.3847/1538-4357/accb59
\bibitem[Jethwa et al.(2020)]{Jet20} Jethwa P., Thater S., Maindl T., Van de Ven G., 2020, ascl.soft. ascl:2011.007
\bibitem[Jin et al.(2020)]{Jin20} Jin, Y., Zhu, L., Long, R. J., et al. 2020, MNRAS, 491, 1690, doi: 10.1093/mnras/stz3072
\bibitem[Jin et al.(2024)]{Jin24} Jin Y., Zhu L., Zibetti S., Costantin L., van de Ven G., Mao S., 2024, A\&A, 681, A95. doi:10.1051/0004-6361/202347197

\bibitem[Kauffmann et~al.(2003b)]{Kau03}Kauffmann, G., Heckman, T. M., Tremonti, C., et al. 2003, MNRAS, 346, 1055
\bibitem[Kewley et~al.(2001)]{Kew01}Kewley, L. J., Dopita, M. A., Sutherland, R. S., Heisler, C. A., Trevena, J., 2001, ApJ, 556, 121
\bibitem[Kennicutt, Tamblyn, \& Congdon(1994)]{ken} Kennicutt R.~C., Tamblyn P., Congdon C.~E., 1994, ApJ, 435, 22. doi:10.1086/174790
\bibitem[Krajnovi{\'c} et al.(2006)]{pafit} Krajnovi{\'c} D., Cappellari M., de Zeeuw P.~T., Copin Y., 2006, MNRAS, 366, 787
\bibitem[Krajnovi{\'c} et al.(2009)]{Kra09} Krajnovi{\'c} D., McDermid R.~M., Cappellari M., Davies R.~L., 2009, MNRAS, 399, 1839. doi:10.1111/j.1365-2966.2009.15415.x
\bibitem[Krajnovi{\'c} et al.(2015)]{Kra15} Krajnovi{\'c}, D., Weilbacher, P. M., Urrutia, T., et al. 2015, MNRAS, 452, 2, doi: 10.1093/mnras/stv958
\bibitem[Kormendy \& Kennicutt(2004)]{KorKen04} Kormendy, J. \& Kennicutt Jr, R., 2004, ARA\&A, 42, 603
\bibitem[Larson(1974)]{Lar74} Larson, R., 1974, MNRAS, 166, 585
\bibitem[Liepold, et al.(2020)]{Lie20} Liepold, C. M., Quenneville, M. E., Ma, C.-P., et al. 2020, ApJ, 891, 4, doi: 10.3847/1538-4357/ab6f71
\bibitem[Lipka \& Thomas(2021)]{LipTho21} Lipka, M., \& Thomas, J. 2021, MNRAS, 504, 4599, doi: 10.1093/mnras/stab1092
\bibitem[Lawrence et al.(2007)]{ukidss} Lawrence A., Warren S.~J., Almaini O., Edge A.~C., Hambly N.~C., Jameson R.~F., Lucas P., et al., 2007, MNRAS, 379, 1599. doi:10.1111/j.1365-2966.2007.12040.x
\bibitem[Lyubenova et al.(2013)]{Lyu13} Lyubenova, M., van den Bosch, R. C. E., C\^ot\'e, P., et al. 2013, MNRAS, 431, 3364, doi: 10.1093/mnras/stt414
\bibitem[Marigo et al.(2013)]{Mar13}Marigo, P. et al. 2013, MNRAS, 434, 488
\bibitem[McMahon et al.(2013)]{vhs} McMahon R.~G., Banerji M., Gonzalez E., Koposov S.~E., Bejar V.~J., Lodieu N., Rebolo R., et al., 2013, Msngr, 154, 35
\bibitem[Miranda et al.(2025)]{Mir25} Miranda H., Pappalardo C., Afonso J., Papaderos P., Lobo C., Paulino-Afonso A., Carvajal R., et al., 2025, A\&A, 694, A102. doi:10.1051/0004-6361/202451648
\bibitem[Neureiter et al.(2021)]{Neu21} Neureiter, B., Thomas, J., Saglia, R., et al. 2021, MNRAS, 500, 1437, doi: 10.1093/mnras/staa3014
\bibitem[Padovani et al.(2017)]{Pad17} Padovani P., Alexander D.~M., Assef R.~J., De Marco B., Giommi P., Hickox R.~C., Richards G.~T., et al., 2017, A\&ARv, 25, 2
\bibitem[Papaderos et al.(2022)]{Pap22} Papaderos P., Breda I., Humphrey A., Michel Gomes J., Ziegler B.~L., Pappalardo C., 2022, A\&A, 658, A74
\bibitem[Plat, et al.(2019)]{Plat19} Plat A., Charlot S., Bruzual G., Feltre A., Vidal-García A., Morisset C., Chevallard J., Todt H., 2019, MNRAS, 490, 978
\bibitem[Poci, et al.(2019)]{Poci19} Poci, A., McDermid, R. M., Zhu, L., \& van de Ven, G. 2019, MNRAS, 487, 3776, doi: 10.1093/mnras/stz1154
\bibitem[Quenneville, et al.(2021)]{Que21} Quenneville, M. E., Liepold, C. M., \& Ma, C.-P. 2021, ApJS, 254, 25, doi: 10.3847/1538-4365/abe6a0
\bibitem[Quilley \& de Lapparent(2023)]{Qui23} Quilley L., de Lapparent V., 2023, A\&A, 680, A49. doi:10.1051/0004-6361/202346774

\bibitem[Rigamonti et al.(2023a)]{bang1} Rigamonti F., Dotti M., Covino S., Haardt F., Landoni M., Del Pozzo W., Lupi A., Zibetti S., 2022a, BANG: BAyesian decomposiotioN of Galaxies, Astrophysics Source Code Library, record ascl: 2205.022
\bibitem[Rigamonti et al.(2023b)]{bang2} Rigamonti F., Dotti M., Covino S., Haardt F., Cortese L., Landoni M., Varisco L., 2023, MNRAS, 525, 1008. doi:10.1093/mnras/stad2363
\bibitem[S\'anchez et al.(2012)]{califa}S\'anchez, S. F., Kennicutt, R. C., Gil de Paz, A., et al. 2012, A\&A, 538, 8
\bibitem[S{\'a}nchez et al.(2016b)]{pipe3d} S{\'a}nchez S.~F., P{\'e}rez E., S{\'a}nchez-Bl{\'a}zquez P., Garc{\'\i}a-Benito R., Ibarra-Mede H.~J., Gonz{\'a}lez J.~J., Rosales-Ortega F.~F., et al., 2016, RMxAA, 52, 171. doi:10.48550/arXiv.1602.01830
\bibitem[S{\'a}nchez et al.(2016c)]{fit3d} S{\'a}nchez S.~F., P{\'e}rez E., S{\'a}nchez-Bl{\'a}zquez P., Gonz{\'a}lez J.~J., Ros{\'a}les-Ortega F.~F., Cano-D{\'\i}az M., L{\'o}pez-Cob{\'a} C., et al., 2016, RMxAA, 52, 21. doi:10.48550/arXiv.1509.08552
\bibitem[de S{\'a}-Freitas et al.(2023)]{SaFre23} de S{\'a}-Freitas C., Gadotti D.~A., Fragkoudi F., Coccato L., Coelho P., de Lorenzo-C{\'a}ceres A., Falc{\'o}n-Barroso J., et al., 2023, A\&A, 678, A202. doi:10.1051/0004-6361/202347028
\bibitem[Salpeter(1955)]{Sal55} Salpeter E.~E., 1955, ApJ, 121, 161. doi:10.1086/145971
\bibitem[Santucci et al.(2022)]{San22} Santucci G., Brough S., van de Sande J., McDermid R.~M., van de Ven G., Zhu L., D'Eugenio F., et al., 2022, ApJ, 930, 153
\bibitem[Santucci et al.(2024)]{San24} Santucci G., Lagos C.~D.~P., Harborne K.~E., Ludlow A., Proctor K.~L., Foster C., McDermid R., et al., 2024, MNRAS, 528, 2326. doi:10.1093/mnras/stae113
\bibitem[Schwarzschild(1979)]{Sch79} Schwarzschild, M. 1979, ApJ, 232, 236, doi: 10.1086/157282
\bibitem[{{Schawinski} {et~al.}(2007){Schawinski}, {Thomas}, {Sarzi}, {Maraston}, {Kaviraj}, {Joo}, {Yi}, \& {Silk}}]{Sch07} {Schawinski}, K., {Thomas}, D., {Sarzi}, M., {et~al.} 2007, \mnras, 382, 1415
\bibitem[Slater et al.(2019)]{Sla19} Slater R., Nagar N.~M., Schnorr-M{\"u}ller A., Storchi-Bergmann T., Finlez C., Lena D., Ramakrishnan V., et al., 2019, A\&A, 621, A83. doi:10.1051/0004-6361/201730634
\bibitem[S\'{e}rsic(1963)]{Sersic63}S\'{e}rsic, J. L., 1963, Boletin de la Asociacion Argentina de Astronomia, vol.6, p.41
\bibitem[Smaji{\'c} et al.(2015)]{Sma15} Smaji{\'c} S., Moser L., Eckart A., Busch G., Combes F., Garc{\'\i}a-Burillo S., Valencia-S.~M., et al., 2015, A\&A, 583, A104. doi:10.1051/0004-6361/201424850
\bibitem[Springel \& Hernquist(2005)]{SprHer05} Springel V. \& Hernquist L., 2005, ApJ, 622, L9
\bibitem[Thater et al.(2017)]{Tha17} Thater, S., Krajnovi{\'c}, D., Bourne, M. A., et al. 2017, A\&A, 597, A18, doi: 10.1051/0004-6361/201629480
\bibitem[Thater et al.(2019)]{Tha19} Thater, S., Krajnovi{\'c}, D., Cappellari, M., et al. 2019, A\&A, 625, A62, doi: 10.1051/0004-6361/201834808
\bibitem[Thater et al.(2022)]{dyn} Thater S., Jethwa P., Tahmasebzadeh B., Zhu L., den Brok M., Santucci G., Ding Y., et al., 2022, A\&A, 667, A51. doi:10.1051/0004-6361/202243926
\bibitem[Thater et al.(2023)]{Tha23} Thater S., Lyubenova M., Fahrion K., Mart{\'\i}n-Navarro I., Jethwa P., Nguyen D.~D., van de Ven G., 2023, A\&A, 675, A18. doi:10.1051/0004-6361/202245362
\bibitem[Valdes et al.(2004)]{IUS} Valdes F., Gupta R., Rose J.~A., Singh H.~P., Bell D.~J., 2004, ApJS, 152, 251. doi:10.1086/386343
\bibitem[van de Ven et al.(2006)]{Van06} van de Ven, G., van den Bosch, R. C. E., Verolme, E. K., \& de Zeeuw, P. T. 2006, A\&A, 445, 513, doi: 10.1051/0004-6361:20053061
\bibitem[van den Bosch et al.(2008)]{VanBos08} van den Bosch R.~C.~E., van de Ven G., Verolme E.~K., Cappellari M., de Zeeuw P.~T., 2008, MNRAS, 385, 647
\bibitem[van Rossum(2020)]{pkl} Van Rossum, G. (2020), The Python Library Reference, release 3.8.2, Python Software Foundation
\bibitem[Vasiliev \& Athanassoula(2015)]{VasAth15} Vasiliev, E., \& Athanassoula, E. 2015, MNRAS, 450, 2842, doi: 10.1093/mnras/stv805
\bibitem[Vasiliev \& Valluri(2020)]{VasVal20} Vasiliev, E., \& Valluri, M. 2020, ApJ, 889, 39, doi: 10.3847/1538-4357/ab5fe0
\bibitem[Vazdekis et al.(2010)]{MILES_SSP} Vazdekis A., S{\'a}nchez-Bl{\'a}zquez P., Falc{\'o}n-Barroso J., Cenarro A.~J., Beasley M.~A., Cardiel N., Gorgas J., et al., 2010, IAUS, 262, 65. doi:10.1017/S174392131000253X
\bibitem[Zeilinger et al.(2001)]{Zei01} Zeilinger W.~W., Vega Beltr{\'a}n J.~C., Rozas M., Beckman J.~E., Pizzella A., Corsini E.~M., Bertola F., 2001, Ap\&SS, 276, 643. doi:10.1023/A:1017548101623
\bibitem[Zotos \& Jung(2018)]{Zot18} Zotos E.~E., Jung C., 2018, MNRAS, 473, 806. doi:10.1093/mnras/stx2398
\bibitem[Zhu et al.(2018b)]{Zhu18b} Zhu L., van den Bosch R., van de Ven G., Lyubenova M., Falc{\'o}n-Barroso J., Meidt S. E., Martig M., et al., 2018b, MNRAS, 473, 3000, doi: 10.1093/mnras/stx2409
\bibitem[Zhu et al.(2018c)]{Zhu18c} Zhu, L., van de Ven, G., van den Bosch, R., et al. 2018c, Nature Astronomy, 2, 233, doi: 10.1038/s41550-017-0348-1
\bibitem[Zhu et al.(2020)]{Zhu20} Zhu L., van de Ven G., Leaman R., Grand R.~J.~J., Falc{\'o}n-Barroso J., Jethwa P., Watkins L.~L., et al., 2020, MNRAS, 496, 1579. doi:10.1093/mnras/staa1584
\end{thebibliography}
